\documentclass[twocolumn,tighten,times,trackchanges]{aastex63}
\usepackage{multirow,bigdelim,amsmath,comment,soul}

\received{-}
\revised{-}
\accepted{-}

\shorttitle{New X-ray source in GW\,170817}
\shortauthors{Hajela et al.}

\begin{document}

\title{Evidence for X-ray emission in excess to the  jet afterglow decay 3.5 yrs after the  binary neutron star merger GW\,170817: a new emission component}

\correspondingauthor{A. Hajela}
\email{ahajela@u.northwestern.edu}

\author{A. ~Hajela}
\affiliation{Center for Interdisciplinary Exploration and Research in Astrophysics and Department of Physics and Astronomy, Northwestern University,\\
2145 Sheridan Road, Evanston, IL 60208-3112, USA}
\author{R. ~Margutti}
\affiliation{Department of Astronomy, University of California, Berkeley, CA 94720-3411, USA}
\author{J. ~S. ~Bright}
\affiliation{Department of Astronomy, University of California, Berkeley, CA 94720-3411, USA}
\author{K. ~D. ~Alexander}
\altaffiliation{NHFP Einstein Fellow}
\affiliation{Center for Interdisciplinary Exploration and Research in Astrophysics and Department of Physics and Astronomy, Northwestern University,\\
2145 Sheridan Road, Evanston, IL 60208-3112, USA}
\author{B. ~D. ~Metzger}
\affiliation{Department of Physics, Columbia University, New York, NY 10027, USA}
\affiliation{Center for Computational Astrophysics, Flatiron Institute, 162 W. 5th Avenue, New York, NY 10011, USA}
\author{V. ~Nedora}
\affiliation{Max Planck Institute for Gravitational Physics (Albert Einstein Institute), Am M{\"u}hlenberg 1, Potsdam 14476, Germany}
\affiliation{Institute for Physics and Astronomy, University of Potsdam, Potsdam 14476, Germany}
\author{A. ~Kathirgamaraju}
\affiliation{Division of Physics, Mathematics and Astronomy, California Institute of Technology, Pasadena, CA 91125, USA}
\author{B. ~Margalit}
\affiliation{Astronomy Department and Theoretical Astrophysics Center, University of California, Berkeley, Berkeley, CA 94720, USA}
\author{D. ~Radice}
\affiliation{Institute for Gravitation and the Cosmos, The Pennsylvania State University, University Park, PA 16802}
\affiliation{Department of Physics, The Pennsylvania State University, University Park, PA 16802}
\affiliation{Department of Astronomy \& Astrophysics, The Pennsylvania State University, University Park, PA 16802}
\author{C. ~Guidorzi}
\affiliation{Department of Physics and Earth Science, University of Ferrara, via Saragat 1, I–44122, Ferrara, Italy}
\affiliation{INFN – Sezione di Ferrara, via Saragat 1, I–44122, Ferrara, Italy}
\affiliation{INAF – Osservatorio di Astrofisica e Scienza dello Spazio di Bologna, Via Piero Gobetti 101, I-40129 Bologna, Italy}
\author{E. ~Berger}
\affiliation{Center for Astrophysics \textbar{} Harvard \& Smithsonian, 60 Garden St., Cambridge, MA 02138, USA}
\author{A. ~MacFadyen}
\affiliation{Center for Cosmology and Particle Physics, Physics Department, New York University, New York, NY 10003, USA}
\author{D. ~Giannios}
\affiliation{Department of Physics and Astronomy, Purdue University, West Lafayette, IN 47907, USA}
\author{R. ~Chornock}
\affiliation{Department of Astronomy, University of California, Berkeley, CA 94720-3411, USA}
\author{I. ~Heywood}
\affiliation{Astrophysics, Department of Physics, University of Oxford, Keble Road, Oxford, OX1 3RH, UK}
\affiliation{Department of Physics and Electronics, Rhodes University, PO Box 94, Makhanda, 6140, South Africa}
\affiliation{South African Radio Astronomy Observatory, 2 Fir Street, Black River Park, Observatory, Cape Town, 7925, South Africa}
\author{L. ~Sironi}
\affiliation{Department of Astronomy and Columbia Astrophysics Laboratory, Columbia University, 550 West 120th Street New York, NY 10027, USA}
\author{O. ~Gottlieb}
\affiliation{Center for Interdisciplinary Exploration and Research in Astrophysics and Department of Physics and Astronomy, Northwestern University,\\
2145 Sheridan Road, Evanston, IL 60208-3112, USA}
\author{D. ~Coppejans}
\affiliation{Department of Physics, University of Warwick, Gibbet Hill Road, Coventry CV4 7AL, UK}
\author{T. ~Laskar}
\affiliation{Department of Astrophysics/IMAPP, Radboud University, PO Box 9010, 6500 GL, The Netherlands}
\author{Y. ~Cendes}
\affiliation{Center for Astrophysics \textbar{} Harvard \& Smithsonian, 60 Garden St., Cambridge, MA 02138, USA}
\author{R. ~Barniol Duran}
\affiliation{Department of Physics and Astronomy, California State University, Sacramento, 6000 J Street, Sacramento, CA 95819-6041, USA}
\author{T. ~Eftekhari} 
\affiliation{Center for Interdisciplinary Exploration and Research in Astrophysics and Department of Physics and Astronomy, Northwestern University,\\
2145 Sheridan Road, Evanston, IL 60208-3112, USA}
\author{W. ~Fong}
\affiliation{Center for Interdisciplinary Exploration and Research in Astrophysics and Department of Physics and Astronomy, Northwestern University,\\
2145 Sheridan Road, Evanston, IL 60208-3112, USA}
\author{A. ~McDowell}
\affiliation{Center for Cosmology and Particle Physics, Physics Department, New York University, New York, NY 10003, USA}
\author{M. ~Nicholl}
\affiliation{School of Physics and Astronomy, University of Birmingham, Birmingham B15 2TT, UK}
\author{X. ~Xie}
\affiliation{Mathematical Sciences and STAG Research Centre, University of Southampton, Southampton SO17 1BJ, UK}
\author{J. ~Zrake}
\affiliation{Clemson University, 118 Kinard Laboratory Clemson, SC 29634, USA}
\author{S. ~Bernuzzi}
\affiliation{Theoretisch-Physikalisches Institut, Friedrich-SchillerUniversit\"{a}t Jena, 07743, Jena, Germany}
\author{F. ~S. ~Broekgaarden}
\affiliation{Center for Astrophysics \textbar{} Harvard \& Smithsonian, 60 Garden St., Cambridge, MA 02138, USA}
\author{C. ~D. ~Kilpatrick}
\affiliation{Center for Interdisciplinary Exploration and Research in Astrophysics and Department of Physics and Astronomy, Northwestern University,\\
2145 Sheridan Road, Evanston, IL 60208-3112, USA}
\author{G. ~Terreran}
\affiliation{Las Cumbres Observatory, 6740 Cortona Dr. Suite 102, Goleta, CA, 93117, USA}
\affiliation{Department of Physics, University of California, Santa Barbara, Santa Barbara, CA, 93106, USA}
\author{V. ~A. ~Villar}
\affiliation{Institute for Gravitation and the Cosmos, The Pennsylvania State University, University Park, PA 16802, USA}
\affiliation{Department of Astronomy \& Astrophysics, The Pennsylvania State University, University Park, PA 16802, USA}
\affiliation{Institute for Computational \& Data Sciences, The Pennsylvania State University, University Park, PA, USA}
\author{P. ~K. ~Blanchard}
\affiliation{Center for Interdisciplinary Exploration and Research in Astrophysics and Department of Physics and Astronomy, Northwestern University,\\
2145 Sheridan Road, Evanston, IL 60208-3112, USA}
\author{S. ~Gomez}
\affiliation{Space Telescope Science Institute, 3700 San Martin Dr, Baltimore, MD 21218, USA}
\author{G. ~Hosseinzadeh}
\affiliation{Steward Observatory, University of Arizona, 933 North Cherry Avenue, Tucson, AZ 85721-0065, USA}
\author{D. ~J. ~Matthews}
\affiliation{Department of Astronomy, University of California, Berkeley, CA 94720-3411, USA}
\author{J. ~C. ~Rastinejad}
\affiliation{Center for Interdisciplinary Exploration and Research in Astrophysics and Department of Physics and Astronomy, Northwestern University,\\
2145 Sheridan Road, Evanston, IL 60208-3112, USA}

\begin{abstract}
For the first $\sim3$ years after the binary neutron star merger event GW\,170817 the  
radio and X-ray radiation has been dominated by emission from a structured relativistic off-axis jet 
propagating into a low-density medium with n $< 0.01\,\rm{cm^{-3}}$.  
We report on  observational evidence for {an excess of X-ray emission} 
at $\delta t>900$ days after the  merger.  
With  
$L_x\approx5\times 10^{38}\,\rm{erg\,s^{-1}}$  at 1234 days, {the recently detected X-ray emission represents a  $\ge 3.2\,\sigma$ (Gaussian equivalent)
deviation from the universal post jet-break model that  best fits the  multi-wavelength afterglow 
at earlier times. In the context of \texttt{JetFit} afterglow models, current data represent a departure with statistical significance $\ge 3.1\,\sigma$, depending on the fireball collimation, with the most realistic models showing excesses at the level of $\ge 3.7\,\sigma$.
} A lack of detectable 3\,GHz radio emission 
suggests 
a harder broad-band spectrum than
the jet afterglow. {These properties are consistent with the emergence of a new emission component such as
synchrotron radiation}  from a mildly relativistic shock generated by the expanding merger ejecta, i.e. a kilonova afterglow.
In this context, we present a set of ab-initio numerical-relativity BNS merger simulations that show that an  X-ray excess supports the presence of a high-velocity tail in the merger ejecta, and argues against the prompt collapse of the merger remnant into a black hole. Radiation from accretion processes on the compact-object remnant represents a viable alternative.  
Neither a kilonova afterglow nor accretion-powered emission have been observed before, as detections of BNS mergers at this phase of evolution are unprecedented.  
\end{abstract}

\section{Introduction} \label{Sec:intro}
The binary neutron-star (BNS) merger GW\,170817 is the first celestial object from which both gravitational waves (GWs) and light have been detected \citep{Abbott+2017multimessenger}, enabling unprecedented insight on the pre-merger (GWs) and post-merger (light) physical properties of these phenomena (\citealt{MarguttiChornock2021,Nakar2020review} and references therein). 
GWs from GW\,170817 were detected on 17 August 2017 at 12:41:04 (UT) by Advanced LIGO and Advanced Virgo \citep{gw170817discovery}. The event was rapidly localized to reside in a nearby galaxy at a distance of $40.7\,$Mpc \citep{Cantiello+2018} thanks to the identification of its electromagnetic counterpart across the spectrum ($\gamma$-rays to radio,  \citealt{Abbott+2017multimessenger}). During the first $\sim 70$\,days, the electromagnetic spectrum of GW\,170817 consisted of a combination of thermal emission partially powered by the radioactive decay of heavy chemical elements freshly synthesized in the merger ejecta (i.e. the ``kilonova'') and non-thermal synchrotron emission dominating in the X-ray and radio bands.  The spectrum and flux evolution of the kilonova emission from GW\,170817 was in agreement with theoretical predictions \citep{Metzger+2017kilonova} demonstrating that mergers of neutron stars are one of the major sources of heavy elements in our Universe (e.g., \citealt{Rosswog+2018cosmicnucl}). Modeling of the UV-Optical-NIR thermal emission from the kilonova allowed estimates of the bulk velocities and masses of the slower-moving ejecta powering the kilonova: $v \sim 0.1$ -- $0.3c$ and total ejecta mass $M_{\rm{ej}} \sim 0.06\,\rm{M_{\odot}}$, carrying a kinetic energy of $\approx 10^{51}\,\rm{erg}$ \citep{Villar+2017combined,Cowperthwaite+2017EM,Drout+2017LC,Kilpatrick+2017,Arcavi_2018,Waxman+2018,Bulla+2019,Nicholl+2021}. 

In the first $\approx 900$ days since merger, the non-thermal spectrum of GW\,170817 has been dominated by synchrotron emission from an ultra-relativistic structured jet initially pointing $\theta_{\rm obs} \sim 15$ -- $25$ degrees away from our line of sight \citep{Mooley+2018superluminal,Ghirlanda+2019,Hotokezaka+2019,Nathanail+2021}. Radio-to-X-ray data did not show any evidence for spectral evolution across nine orders of magnitude of frequency for 900 days \citep{Fong+2019,Hajela+2019,Troja+2020} 
and the emission was well characterized as originating from an optically thin synchrotron source with a power-law spectrum $F_{\rm \nu}\propto \nu^{-(p-1)/2}$ with best-fitting $p=2.166 \pm 0.026$ \citep{Fong+2019}, where $p$ is the index of the distribution of relativistic electrons responsible for the emission  $dN_{\rm e}/d\gamma_{\rm e}\propto \gamma_{\rm e}^{-p}$, and $\gamma_{\rm e}$ is the electron Lorentz factor above some minimum Lorentz factor $\gamma_{\rm min}$. Modeling  of the multi-wavelength off-axis jet afterglow emission enabled tight constraints on some of the system and environment parameters (or their combination): for example, the jet kinetic energy to environment density ratio was constrained to $E_{\rm{k}}/n\approx (1$ -- $2)\times 10^{53}\,\rm{erg\,cm^{3}}$ \citep{Mooley+2018superluminal,Ghirlanda+2019,Hotokezaka+2019} 
with a credible density range of $10^{-4}\,{\rm{cm^{-3}}}\le n \le 10^{-2}\,\rm{cm^{-3}}$ \citep{Hajela+2019} and the inferred ultra-relativistic jet opening angle is $\theta_{\rm{jet}} \approx 2$ -- $5$ degrees \citep{Mooley+2018superluminal,Ghirlanda+2019,Hotokezaka+2019,Nathanail+2021}. A robust prediction of the off-axis afterglow model post-peak (i.e. after radiation from the core of the jet enters the observer's line of sight) is that of a universal asymptotic light-curve decay with flux $F_{\rm \nu}(t)\propto t^{-p}$ \citep{Sari+1999}. For the best-fitting jet-environment parameters of GW\,170817 no broadband spectral evolution is expected, leading to  $F_{\rm \nu}(\nu,t)\propto \nu^{-(p-1)/2} t^{-p}$ (we call this ``universal post jet-break model"). Until $\approx 900$ days post-merger, panchromatic observations of the jet afterglow of GW\,170817 satisfied these expectations. 

Here we present the results from our multi-wavelength campaign of GW\,170817 at X-ray and radio frequencies extending to $1273$ days since merger, which was designed to constrain the emergence of the kilonova afterglow. These observations provide the first statistically significant evidence for {an excess of X-ray emission that is consistent with the emergence of}
a new X-ray component of emission that is physically distinct from the jet afterglow.  This paper is organized as follows. In \S\ref{Sec:xray}, we present the detailed analysis of the recent X-ray observations taken at $\delta t=1234\,\rm{days}$ and a re-analysis of the observations at $\delta t \sim 939\,\rm{days}$ and earlier. Newly acquired VLA and MeerKAT observations around $\delta t =1234$ days are presented in \S\ref{Sec:radio}. We calculate the statistical evidence for the observed excess of emission in X-rays in \S\ref{Sec:Excess_stats}. We discuss the inferences of the observed excess in X-rays and a lack of detection in radio on the broadband spectrum in \S\ref{Sec:SpecEvol}. We discuss the possibility of different scenarios leading to an excess in X-rays without an accompanying  radio  emission as the observations suggest in \S\ref{Sec:latetimejet}-\S\ref{Sec:Engine}. We draw our conclusions in \S\ref{Sec:summary}.

\begin{figure*}[!t]
\begin{center}
\includegraphics[scale=0.25]{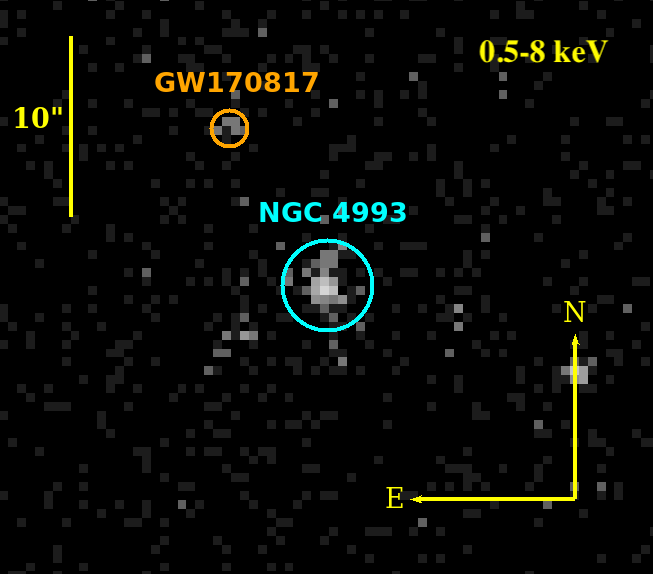}
\includegraphics[scale=0.2515]{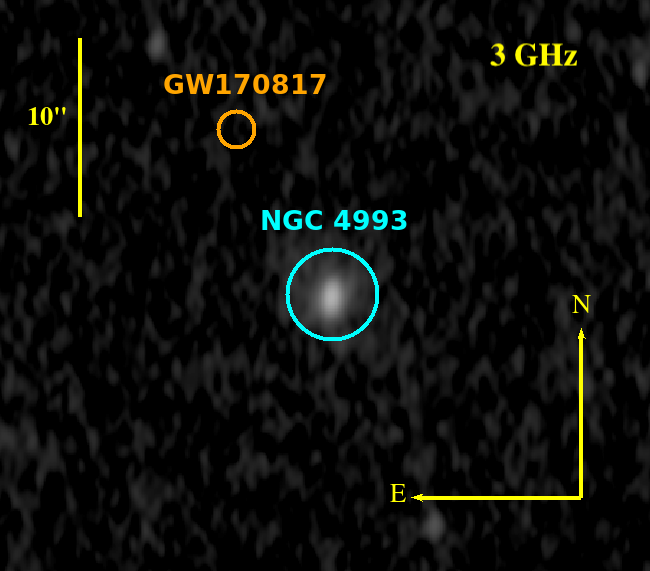}
\caption{
\emph{Left Panel}: Combined X-ray image consisting of \emph{CXO} observations spanning $\delta t= 1209$ -- $1258$ days in the $0.5$ -- $8$\,keV energy range. An X-ray source is clearly detected at the location of GW\,170817 with statistical significance of 7.2$\,\sigma$ ( Table \ref{Tab:xrayanalysistab}).  \emph{Right Panel}: Combined radio image comprising VLA 3\,GHz observations acquired in the time range $\delta t =1216$ -- $1265$ days. No radio emission is detected at the location of GW\,170817. The RMS noise around the location of the BNS merger is $\sim  1.7\,\rm{\mu Jy}$ (\S\ref{SubSec:VLA_obs}).  In both panels the orange and light-blue regions have a 1\arcsec\, and 2.5\arcsec\, radius, respectively, and mark the location of the BNS merger and its host galaxy. }
\label{Fig:xrayradioimg}
\end{center}
\vspace{-0.25cm}
\end{figure*}

\begin{figure*}[!t]
\begin{center}
\includegraphics[scale=0.65]{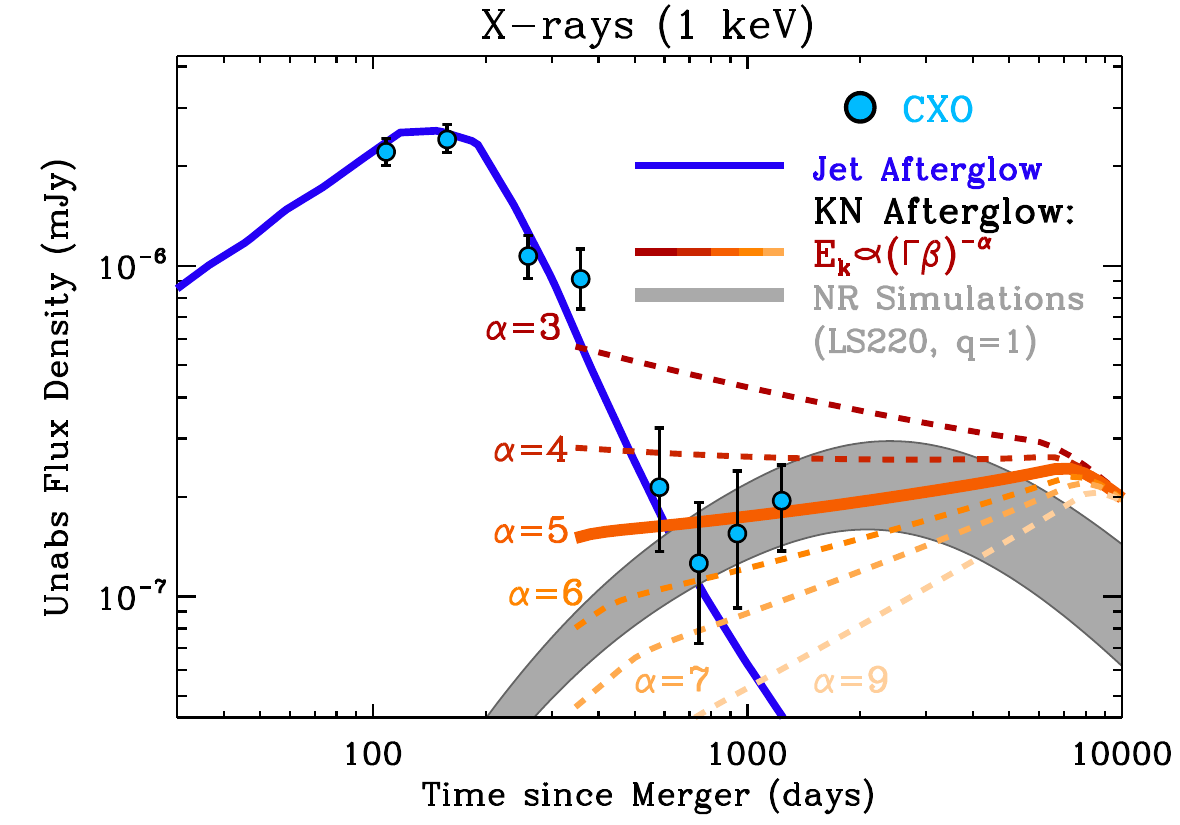}
\includegraphics[scale=0.65]{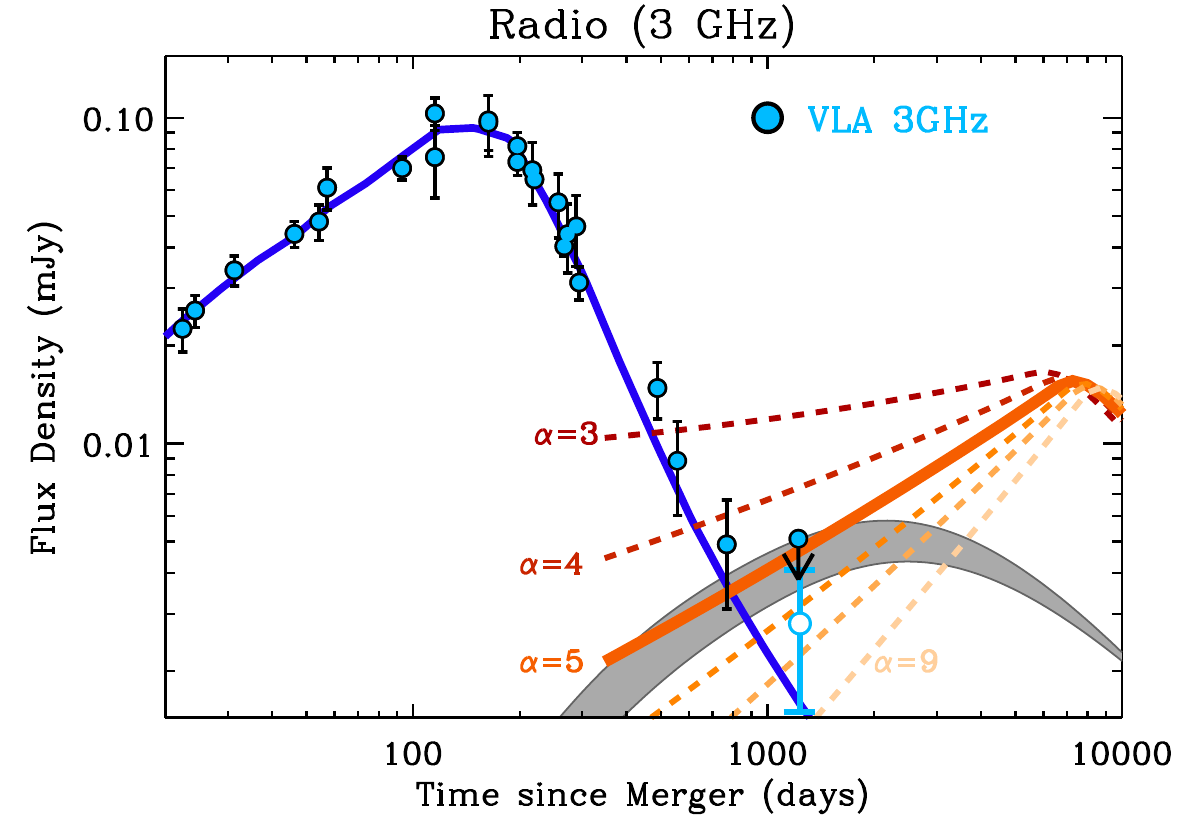}
\caption{
X-ray (\emph{upper panel}) and radio (3 GHz, \emph{lower panel}) evolution of the emission from GW\,170817 as detected by the \emph{CXO} and the VLA (light-blue circles). Open circle: peak pixel flux value within one synthesized beam at the location of GW\,170817 from \citealt{Balasubramanian21}. At $\delta t>900$ days the X-ray emission shows an excess compared to the off-axis jet afterglow model (solid blue line, \S\ref{Sec:Excess_stats} and  \S\ref{Sec:latetimejet}) that indicates the emergence of a new emission component. Red-to-orange dashed lines: synchrotron radiation from the kilonova afterglow calculated using semi-analytical models \citep{Kathirgamaraju+2019}\, where we parametrized the kilonova kinetic energy distribution as $E_{\rm{k}}\propto (\Gamma \beta)^{-\alpha}$ for $\beta \ge 0.35$ and we used a total kilonova kinetic energy of $10^{51}$ erg. These models require $p<2.15$ to avoid violating our radio upper limit. Here we use $p=2.05$ and we emphasize with a solid thick line the $\alpha=5$ model. Other kilonova afterglow parameters assumed: $\epsilon_{\rm{B}}=0.001$, $\epsilon_{\rm{e}}=0.1$, $n=0.001\,\rm{cm^{-3}}$. Grey shaded area: synchrotron emission calculated from kilonova kinetic ejecta profiles derived from ab-initio numerical relativity simulations using a neutron-star mass-ratio $q=1$ and the LS220 equation of state (\S\ref{Sec:KNmodels}). These simulations emphasize the contribution from the merger's dynamical ejecta. The shaded area corresponds to  values $p_{KN}=2.05$ -- $2.15$, $n=6\times 10^{-3}\,\rm{cm^{-3}}$, $\epsilon_{\rm{e}}=0.1$ and $\epsilon_{\rm{B}}=0.01$. }
\label{Fig:Xrays}
\end{center}
\end{figure*}

\section{X-ray Observations}\label{Sec:xray}
We present \emph{Chandra X-ray Observatory} (\emph{CXO}) observations of the X-ray emission from GW\,170817 acquired at $\delta t=1209$ -- $1258$\,days since merger (combined X-ray image in the left panel in Figure 1) and a complete and homogeneous analysis of the entire \emph{CXO} data set. 
Results from \emph{CXO} observations of the jet afterglow of GW\,170817 in the time range $\delta t=2.33$ -- $939.31$ days have already been published in the literature \citep{Margutti+2017electromagnetic, Haggard+2017chandra, Margutti+2018binary, Nynka+2018fadingxray, Alexander+2018, Troja+2018outflow,Ruan+2018xray, Pooley+2018, Piro+2019, Troja+2019,Hajela+2019, Hajela+2020, Troja+2020, Makhathini+2020}.
With respect to previous data reductions: (1) when possible, we do not assume a spectral model for the X-ray count-to-flux calibration, which allows us to test for spectral evolution; (2) we align all the X-ray images to a common astrometric solution, significantly improving on the \emph{CXO} relative astrometry; (3) for each observation we extract a spectrum and we perform a flux calibration that utilizes the complete information on the instrumental response at the time of the observation (as opposed to using averaged instrumental responses); (4) we jointly fit spectra from observations acquired close in time (i.e. around the same ``epoch'') as opposed to merging the files into an average spectrum; (5) we implement an accurate point-spread function (PSF) correction; (6) we calculate the model parameter uncertainties (including the unabsorbed fluxes) with MCMC simulations that self-consistently account for the low-count statistics and the deviation from Gaussian statistics. {At $\delta t>900$ days, the low number statistics of the detected X-rays does not allow us to independently constrain the spectral model and we thus offer a flux calibration that \emph{assumes} the jet-afterglow spectral parameters.}

\subsection{CXO Source Count Rates}\label{SubSec:CXOsrcctrate}
We observed GW\,170817 with the \emph{CXO} from December 09, 2020 at 00:05:21 UT through December 13, 2020 at 14:02:43 UT, and further between January 18, 2021 at 09:43:15 UT and January 27, 2021 at 08:49:13 UT, spanning $\delta t = 1209$ -- $1258$\,days after the merger. The observation was taken in seven distinct exposures (Obs ID 22677, 24887, 24888, 24889, 23870, 24923, and 24924; PI Margutti; programs \#21510449 and \#22510329, publicly available on the \emph{CXO} archive) for a total exposure time of $189.1$\,ks.

We reprocessed the entire  \emph{CXO} dataset  using the \texttt{repro} task within \texttt{CIAO} (v4.13.0, \citealt{Fruscione+2006}) with standard ACIS data filtering and using the latest calibration database (\texttt{CALDB, v4.9.4}). We used \texttt{wcs\_match} and \texttt{wcs\_update} to realign all the IDs to a common astrometric solution using as a reference the list of X-ray point-source positions generated with \texttt{wavdetect} run on our longest exposure observation (Obs ID 20860). In ID 20860 the X-ray emission from GW\,170817 is detected with high significance at sky coordinates RA=13$^h$09$^m$48$^s$.061 $\pm$ 0.049$^s$  
and dec=-23\arcdeg:22\arcmin:52.88\arcsec\, $\pm$ 0.034$\arcsec$\, 
(J2000). 
After having realigned the images, for each ID we extracted source count-rates and spectra using a 1\arcsec\, region centered at the coordinates above.  Table \ref{Tab:xrayanalysistab} lists the inferred $0.5$ -- $8$\,keV net count-rates and the associated targeted-detection significance.  For source detection, we employed a 1\arcsec\, source region and we filtered in the energy range $0.5$ -- $8$\,keV to minimize the background contribution. For reference, a 1\arcsec\, region contains $\gtrsim90$\% of the PSF at 1 keV.

Focusing on the data at $\delta t > 900$\,days, we find that 
 an X-ray source is clearly detected at the location of GW\,170817 at $\delta t = 939$\,days with a statistical significance of $5.4\,\sigma$ (Gaussian equivalent), 
corresponding to a net count-rate of (7.53 $\pm$ 2.93)$\times 10^{-5}\,\rm{ct\,s^{-1}}$ ($0.5$ -- $8$\,keV). 
For  \emph{CXO} observations acquired at
$\delta t = 1209$ -- $1214$\,days, we infer  an observed net count-rate of (1.13 $\pm$ 0.36)$\times 10^{-4}\,\rm{ct\,s^{-1}}$ 
($6.3\,\sigma$ detection significance), whereas for the remaining observations acquired between $\delta t = 1250$ -- $1258$\,days, the observed net count-rate is (4.31 $\pm$ 2.28)$\times 10^{-5}\,\rm{ct\,s^{-1}}$ and an X-ray source is detected at a significance level of $3.4\,\sigma$.  
Being temporally close, we combined the latter two sets of observations spanning $1205$ -- $1258$ days and we infer a net count-rate of (7.68 $\pm$ 2.12)$\times 10^{-5}\,\rm{ct\,s^{-1}}$, where an X-ray source is detected with a $7.2\,\sigma$ statistical significance.

\subsection{CXO Spectral Analysis}\label{SubSec:CXOspecanalysis}For each re-aligned Obs ID  we extracted 
a spectrum using a 1\arcsec\, circular source region centered at the location of the X-ray counterpart of GW\,170817 indicated above and a source-free background region of 22\arcsec. 
We used 
\texttt{specextract}, setting the \texttt{refcoord} parameter to the center of the source region to ensure an accurate 
PSF correction to the inferred fluxes. This procedure furthermore ensures that the appropriate instrumental  ARF (Auxiliary Response File) and RMF (Redistribution Matrix File) response files are generated for each Obs ID. We note that not setting \texttt{refcoord} parameter explicitly leads to an overestimate of the PSF correction by an average factor of $\approx 1.2-1.5$ for a source region of 1\arcsec. 
We fitted the data with an absorbed power-law spectral model
(\texttt{tbabs*ztbabs*cflux(pow)} within \emph{Xspec} (v12.9.1). We adopted a Galactic neutral hydrogen column density in the direction of GW\,170817 of NH$_{\rm gal} = 7.84 \times 10^{20}\,\rm{cm^{-2}}$ \citep{Kalberla+2005}. Consistent with results from previous analysis \citep{Margutti+2017electromagnetic,Margutti+2018binary,Alexander+2018,Hajela+2019,Makhathini+2020},  
we did not find evidence for intrinsic absorption and we thus assumed no intrinsic absorption in the following analysis. For $\delta t<750$\,days, we jointly fitted the observations acquired around the same epoch since merger leaving the spectral photon index, $\Gamma$, and the unabsorbed $0.3$ -- $10$\,keV flux as free parameters. We fitted the data in the $0.3$ -- $10$\,keV energy range. We note that filtering the data in the $0.5$ -- $8$\,keV energy range before fitting does not lead to significantly different inferences. We used Cash statistics and we employed a chain of $10^5$ MCMC simulations to estimate the parameter uncertainties to account for the deviation from Gaussian statistics in the regime of low counts. The results from our X-ray spectral modeling are reported in  Table \ref{Tab:xrayanalysistab}. We find no evidence for X-ray spectral evolution of the source at  $\delta t < 745$\,days.
From a joint spectral fit of all \emph{CXO} observations at $\delta t < 745$\,days with the same $\Gamma$ we infer a best-fitting $\Gamma=1.603_{-0.076}^{+0.102}$, consistent with our previous analysis of these observations in \citealt{Hajela+2019} which used a previous \texttt{CALDB} v4.8.3 and a $1.5$\arcsec\, source region.

We now consider the \emph{CXO} observations acquired at $\delta t > 745$\,days. These \emph{CXO} observations were acquired in two epochs at $\delta t = 939$ and $\delta t  = 1234$\,days since merger. 
The low-count statistics of   $6$ and $12$ photons, respectively, available for model fitting after \emph{Xspec} filtering in the $0.3$ -- $10$\,keV energy range leads to poorly constrained spectral photon indexes $\Gamma=1.16^{+1.38}_{-1.39}$ and $\Gamma=1.92^{+2.53}_{-0.65}$. We thus proceeded by freezing the spectral photon index to $\Gamma = 1.603$ (i.e. the best-fit value inferred from the joint fit of all the \emph{CXO} data collected at $\delta t < 745\,\rm{days}$) for the purpose of count-to-flux calibration. The inferred unabsorbed $0.3$ -- $10$\,keV flux is $F_{\rm x}=1.81_{-0.94}^{+0.79}\times10^{-15}\,\rm{erg\,cm^{-2}\,s^{-1}}$  
at $\delta t = 939$ days, and $F_{\rm x}=2.31_{+0.57}^{-0.81}\times 10^{-15}\,\rm{erg\,cm^{-2}\,s^{-1}}$   
at $\delta t =1234$ days, corresponding to luminosities of $L_{\rm x}\approx (3-5)\times 10^{38}\,\rm{erg\,s^{-1}}$ (Table \ref{Tab:xrayanalysistab}).  These recent observations when visually examined against the jet afterglow model that best-fitted the multiwavelength data at $\delta t < 745$\,days (Figure 2) show slight deviations from the expectations. We quantify this deviation in Section 4.

We end by addressing the possibility of X-ray spectral evolution at $\delta t > 745$\,days. We assessed the statistical evidence for X-ray spectral evolution in two ways. First, from a joint spectral modeling of all \emph{CXO} data acquired at $\delta t > 745$\,days with a power-law spectrum, we infer $\Gamma=1.54^{+0.83}_{-0.75}$. 
Compared to $\Gamma=1.603_{-0.076}^{+0.102}$ of the earlier X-ray data reported above, we find that there is no evidence for statistically significant X-ray spectral evolution from this analysis. Second, we generated $10^6$ synthetic spectra of $N=12$ photons (as observed at $\delta t = 1234$\,days in the $0.3$ -- $10$\,keV energy range after \emph{Xspec} filtering) by randomly sampling the probability density distribution associated with an incoming $\Gamma=1.6$ spectrum with NH$_{\rm gal} = 7.84\times 10^{20}\,\rm{cm^{-2}}$ convolved with the \emph{CXO} instrumental response.  We applied the non-parametric distribution-free Epps–Singleton two-sample test to each sample and the parent distribution and we found that $ 52\%$ of the synthetic samples have a p-value at least as extreme as the one associated with the observed photon distribution, leading to no statistical evidence of a departure of the detected photon distribution at $\delta t > 745$\,days from earlier X-ray data.  We conclude that there is no statistically significant evidence for the evolution of the X-ray spectrum at $\delta t > 745$\,days.

Finally, we compare the results from our X-ray analysis with previous results that appeared in the literature, and specifically with the analyses by \citealt{Troja+2020}\, (for $\delta t = 582$ -- $945$\,days), \citealt{Makhathini+2020}\, (for $\delta t = 9$ -- $745$\,days) and \cite{Troja+2021}. 
The analysis by \cite{Makhathini+2020} cannot be used to test for X-ray spectral evolution of the source because the final count-to-flux calibration is performed by assuming a spectral photon index. We find that the central values of the X-ray fluxes reported by \cite{Makhathini+2020} using a 1\arcsec\, source region are systematically larger than our fluxes (by a factor of up to 30\%). Discrepancy remains even after adopting the same $\Gamma= 1.57$ for the count-to-flux calibration.  We are able to reproduce the \cite{Makhathini+2020} X-ray fluxes by removing the \texttt{refcoord} parameter setting from \texttt{specextract}, which leads to artificially inflated PSF corrections of $\approx 20$ -- $50$\%, as previously noted.    Our X-ray fluxes in the time range $\delta t = 582$ -- $945$\,days are  consistent with those reported by \cite{Troja+2020} within 1-$\sigma$ uncertainties. We found that we could reproduce the \cite{Troja+2020}  fluxes by using the online Portable Interactive Multi-Mission Simulator (PIMMS) for the count-rate to flux calibration. In contrast, our spectral analysis and count-to-flux calibration is based on ARFs and RMFs generated from each individual Obs ID to best account for the instrumental response at the time and in the conditions of the observation, as opposed to the proposal planning tool PIMMS.\footnote{https://cxc.harvard.edu/ciao/why/pimms.html}

We finally compare our results with the recent work of \cite{Troja+2021}, which appeared after our first submission to the archive, where their fiducial X-ray fluxes are calculated by adopting a photon index $\Gamma = 1.585$ derived from their modeling of the broadband radio-to-X-ray afterglow spectrum, differently from our jet afterglow \emph{model-independent} analysis. 
The count-to-flux calibration by \cite{Troja+2021} is done by using the hardness ratio (HR) of observed counts in the $0.5$ -- $2$\,keV energy range and the $2$ -- $7$\,keV energy range to infer a spectral photon index, as opposed to the full spectral extraction and analysis that we perform here. The HR method does not account for and does not self-consistently model the uncertainty that affects the energy of each event on the detector. It is furthermore based on averaged instrumental responses (one for each epoch), instead of using the accurate instrumental information from each obs ID, which our joint spectral analysis does. While we emphasize that there is no tension between the derived $0.3$ -- $10$\,keV fluxes of the two methods and the fluxes are consistent to within 1-$\sigma$ uncertainties\footnote{For observations taken at $\delta t = 1234$ day, assuming NH$_{\rm gal}=1.1\times10^{21}\,\rm{cm^{-2}}$ and $\Gamma=1.585$, as in \citet{Troja+2021}, we find an unabsorbed 0.3-10 keV flux of $1.8^{+0.4}_{-0.5} \times 10^{-15}\,\rm{erg\,cm^{-2}\,s}$ when using an averaged instrumental response, which is entirely consistent with $F_x=1.6^{+0.5}_{-0.5}\times 10^{-15}\,\rm{erg\,cm^{-2}\,s}$ reported by \citet{Troja+2021} for the same model parameters. Instead, using the more accurate approach of jointly fitting the observations, each with its own instrumental response, and using the same parameters NH$_{\rm gal}=1.1\times10^{21}\,\rm{cm^{-2}}$ and $\Gamma=1.585$, we find an unabsorbed flux of $2.51^{+0.66}_{-0.92}\times 10^{-15}\,\rm{erg\,cm^{-2}\,s}$, which is consistent to our value. All errors are quoted at 68\% confidence level.} (Figure \ref{Fig:xraycomparison}), here we note the following: (i) as there is no evidence of the latest radio and X-ray observations at $\delta t = 1234$\,days lying on the same power-law segment of the spectrum, no such assumption is made for the flux calibration of the X-ray data. 
(ii) We self-consistently propagate the uncertainties during the spectral calibration.
(iii) The X-ray upper limit at $2$ days is computed using pure Poisson statistics and represents the 3$\sigma$ deviation from the background, as we detail in \cite{Margutti+2017electromagnetic}. The difference in the flux limit is partially a result of the different photon indexes assumed by \cite{Troja+2021} ($\Gamma=1.585$) and \cite{Margutti+2017electromagnetic} ($\Gamma=2$) for the spectral calibration of the count-rate upper limit. Unsurprisingly, models with harder assumed photon indices lead to greater $0.3$ -- $10$\,keV fluxes.

\section{Radio Observations}\label{Sec:radio}
\subsection{VLA Data Analysis} \label{SubSec:VLA_obs}
We initiated late-time S and Ku-band Karl G. Jansky Very Large Array (VLA) observations of GW\,170817 as part of our joint \emph{CXO}-VLA proposals \#21510449 and \#22510329 (PI Margutti). GW\,170817 was observed for a total of 10.21 hours  on source at S-band spread between three observations occurring on 15\textsuperscript{th} December 2020 ($\delta t=1216.08$), 27\textsuperscript{th} December 2020 ($\delta t=1228.02$) and 2\textsuperscript{nd} February 2021 ($\delta t=1264.95$). All three observations were conducted while the VLA was in A-configuration and at a central frequency of $3\,\rm{GHz}$ using a $2\,\rm{GHz}$ bandwidth.
Additionally, we conducted a single observation at Ku-band on 10\textsuperscript{th} February 2021 ($\delta t$=1272.88) for a total of 2.74 hours on source. The observation was conducted with the VLA in A-configuration and at a central frequency of $15\,\rm{GHz}$ using a $6\,\rm{GHz}$ bandwidth.
These data are publicly available on the VLA archive under project IDs SL0449 and SM0329.  Details of each observation are given in  Table  \ref{tab:radioobslogtab}.

Each individual observation was independently calibrated using the VLA calibration pipeline version 2020.1.0.36 as part of CASA (v6.1.2.7, \citealt{mcmullin2007}), with 3C286 used as the flux density and bandpass calibrator and J1258-2219 used to calibrate the time-varying complex gains. We then manually inspected and validated the output and re-ran the pipeline after flagging additional radio frequency interference (RFI). Additional RFI flagging was performed on the results of the second pipeline run. In order to achieve maximum sensitivity we combined the three epochs of S-band data into a single measurement set (right panel in Figure 1) using the CASA task \textsc{concat}. We imaged the concatenated data using \texttt{wsclean} \citep{offringa-wsclean-2014,offringa-wsclean-2017},  creating a $16384\times16384$ pixel image with a single pixel corresponding to $0.08$\arcsec. The synthesised beam is $1.19''\times0.66''$ with a position angle of $-5.57$ degrees. In order to account for spectral variation introduced for sources far from the phase center (we are imaging well beyond the half power point of the primary beam in order to ensure complete deconvolution, and to produce an accurate sky model for self-calibration) we fit a third order polynomial (\texttt{fit-spectral-pol 4}) over eight output channels (\texttt{channels-out 64}). No time or frequency averaging was performed when imaging in order to avoid bandwidth or temporal smearing of sources far from the phase center ensuring the best possible deconvolution. We performed one round of phase-only self-calibration using a sky-model produced from our phase reference calibrated data. 

We do not detect any significant emission at the position of GW\,170817. The root-mean-square (RMS) noise at the edge of the image in a region free of sources is $\sim1.2\,\mu\rm{Jy}$ while in a circular region with 25 pixel radius centered on the position of GW\,170817 we measure an RMS noise of $\sim1.7\,\mu\rm{Jy}$.

The single Ku-band epoch was calibrated using the VLA calibration pipeline version 2020.1.0.36 as part of CASA version 6.1.2.7 and validated by NRAO as part of the Science Ready Data Products project. We imaged the calibrated measure set using the CASA task \texttt{tclean} with a user defined mask. We created a $2048\times2048$ pixel image with a cell size of $0.02''$. We do not detect any significant emission at the location of GW\,170817 and measure an RMS noise of $1.7\,\mu\rm{Jy}$ in a $30\times30$ pixel region centred on the position of GW\,170817.

\subsection{MeerKAT Data Analysis} \label{SubSec:MeerKAT_obs}
We conducted a single observation of the field of GW\,170817 with the MeerKAT radio interferometer on the 3\textsuperscript{rd} January 2021 as part of a DDT request (DDT-20201218-JB-01). These data are available publicly on the SARAO archive. Data were recorded for a total of 7.56 hours (resulting in 7.24 hours on source) with an $8\,\rm{s}$ dump time at the UHF band between $544\,\rm{MHz}$ and $1088\,\rm{MHz}$ with a central frequency of $816\,\rm{MHz}$ using 4096 frequency channels. Details of the observation are given in  Table \ref{tab:radioobslogtab}.

Data reduction was performed using \textsc{oxkat} \citep{Heywood+2020}, a suite of semi-automated scripts to reduce MeerKAT UHF and L-band data.  First, phase reference calibration (1GC) is carried out using CASA, with flagging performed with Tricolour (a variant of the SARAO Science Data Processing flagging software). B0407$-$65 was observed to calibrate the flux and bandpass of the instrument and 3C283 was used to calibrate the time variable complex gains. Second, we used \textsc{wsclean} to image the field and the resulting sky model was used to perform phase and delay self-calibration (2GC) using \textsc{cubical} \citep{Kenyon+2018}. Images created throughout this process are $10240 \times 10240$ pixels with a robust weighting of $-$0.5 and pixel size of 1.7\arcsec. The phase calibrator, 3C283, is a bright off-axis source when observing GW\,170817 with MeerKAT at the UHF band and leaves strong imaging artefacts after 2GC calibration. Strong sources away from the phase center of an interferometer have their apparent spectral shape modified by the time and frequency dependent primary beam, and for a sufficiently wide field of view one set of gain solutions (direction-dependent) is not appropriate to properly calibrate the data. These issues result in a corrupted point spread function that will not vanish under deconvolution. The primary beam can be corrected for either by providing a model of the primary beam for the array, or by using higher order polynomials when fitting the spectral variation when cleaning (\textsc{oxkat} employs the latter). To correct for direction-dependent gains across the wide MeerKAT field of view ($\sim2\,\rm{deg}^{2}$ at UHF) we `peel' the source 3C283, and performed faceted direction dependent self-calibration on the residual data. The peeling stage was performed using \textsc{cubical}, and the facet based direction dependent calibration was carried out using \textsc{killms}\, with \textsc{ddfacet} \citep{Tasse+2018} used to image the corrected data. To enhance the resolution we image the final data-set with a Briggs robustness parameter $-1$.

The RMS noise at the edge of the image in a region free of sources is $8.5\,\mu\rm{Jy}$. Due to the extremely high source brightness sensitivity of MeerKAT the region around the phase center has a very high density of sources, making it difficult to estimate the phase center noise. We opt to fit the entire image for significant emission using PyBDSF \citep{Mohan+2015} using island and pixel thresholds of $3\sigma$ and $5\sigma$, respectively, with adaptive RMS thresholding turned on. We identify extended (resolved) emission from the host galaxy of GW\,170817 (NGC\,4993) and emission from a source. We identify no significant emission at the position of GW\,170817. Using a $40\times40$ pixel region centered on the position of GW\,170817 we measure an RMS noise of $\sim13\,\mu\rm{Jy}$.

\section{Statistical Evidence for an X-ray Excess of Emission} \label{Sec:Excess_stats}
\begin{figure*}[!t]
\vskip 0.5 true cm
\begin{minipage}{.5\textwidth}
    \centering
    \includegraphics[scale=0.25,width=\textwidth]{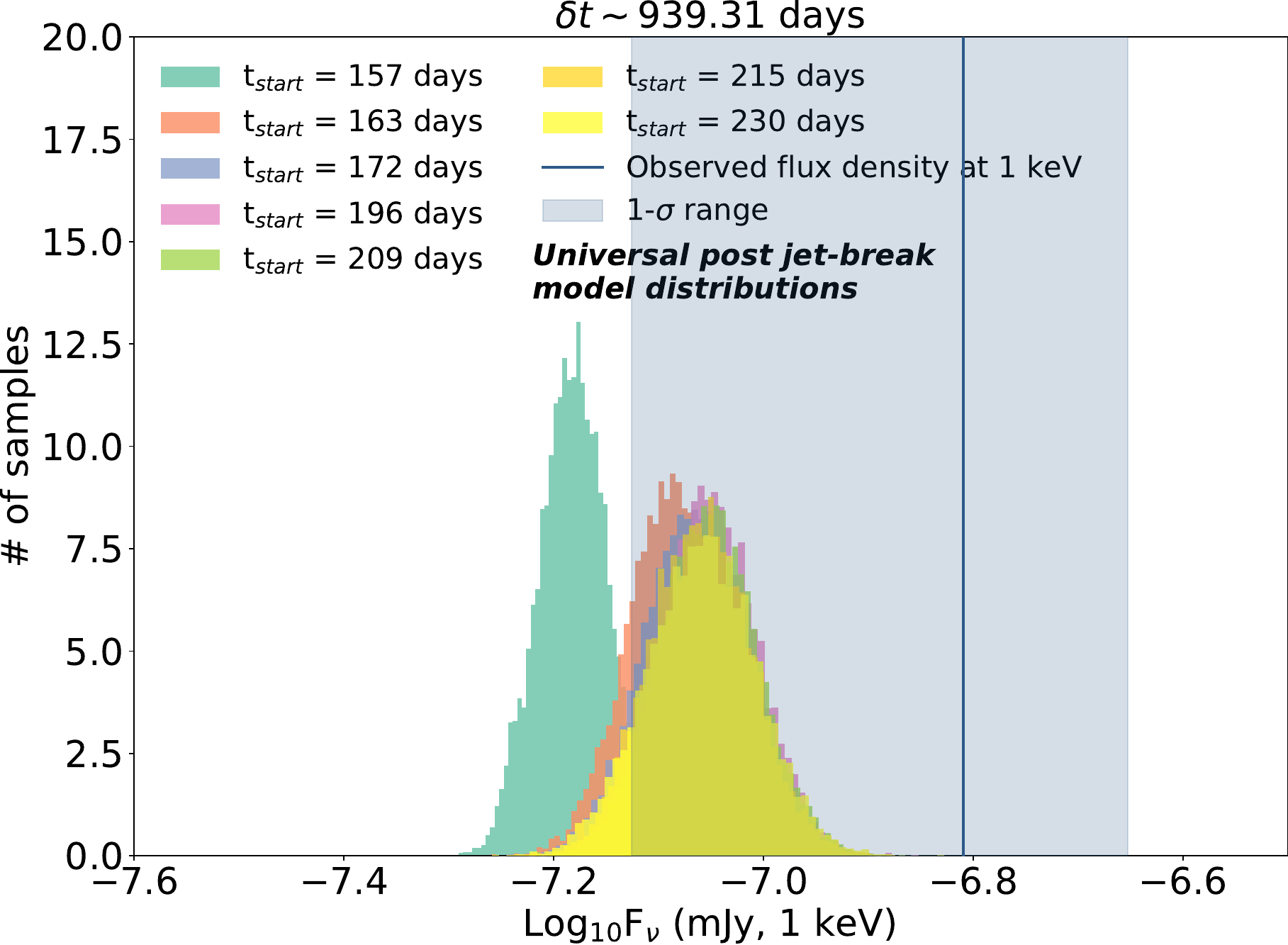}
\end{minipage}
\begin{minipage}{.5\textwidth}
    \centering
    \includegraphics[scale=0.25,width=\textwidth]{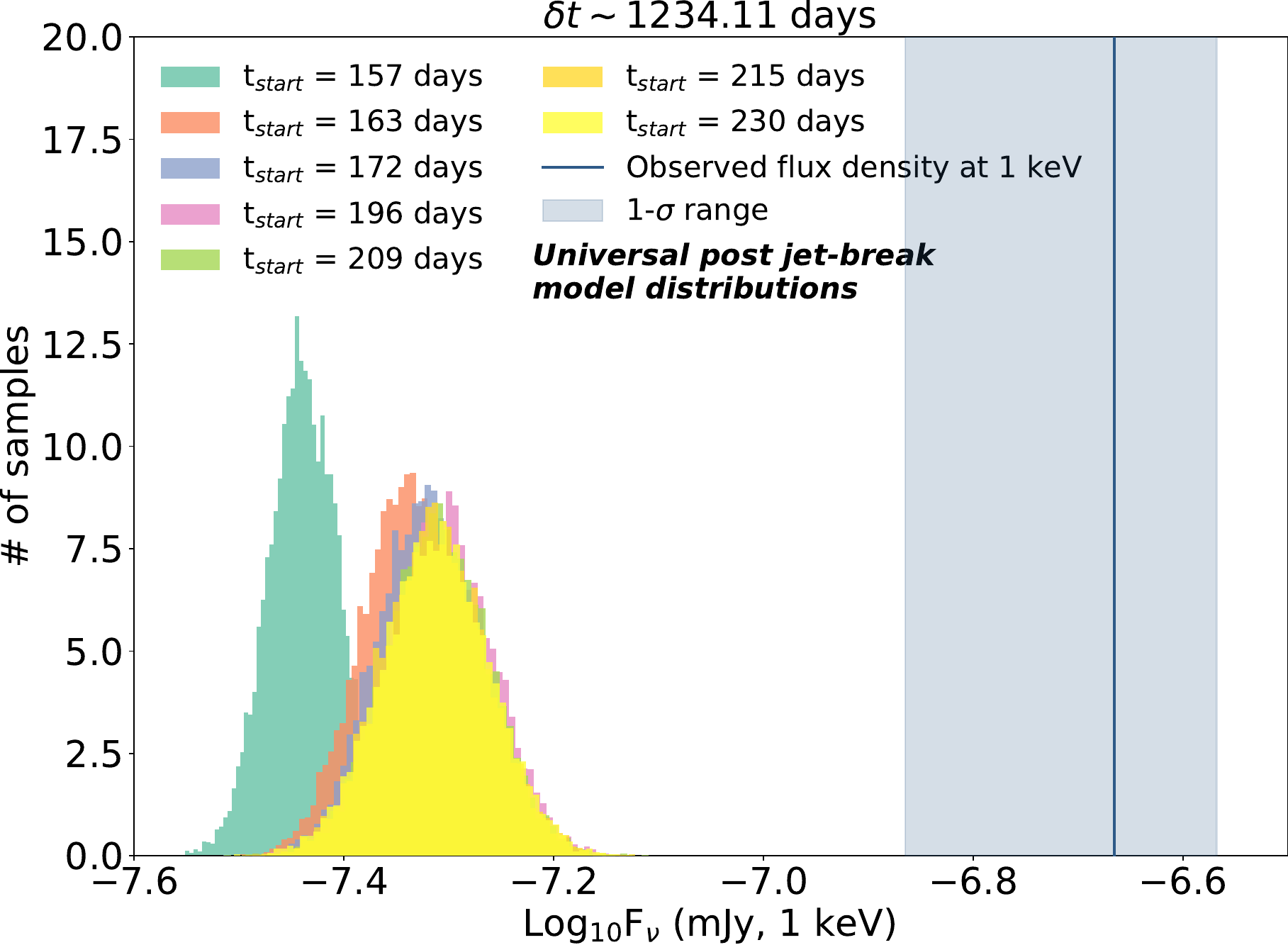}
\end{minipage}
\vskip 1 true cm
\begin{minipage}{.5\textwidth}
    \centering
    \includegraphics[scale=0.25,width=\textwidth]{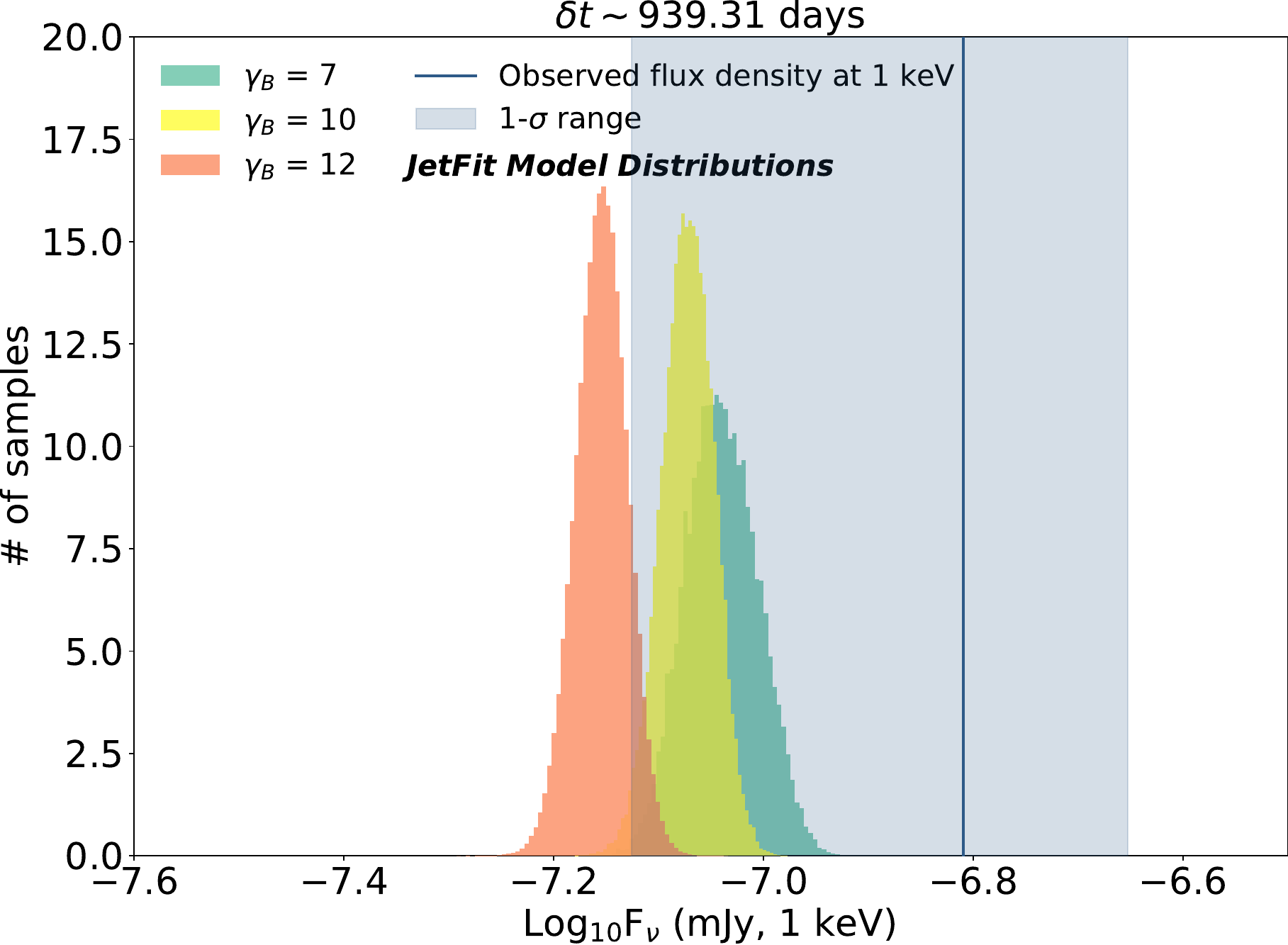}
\end{minipage}
\begin{minipage}{.5\textwidth}
    \centering
    \includegraphics[scale=0.25,width=\textwidth]{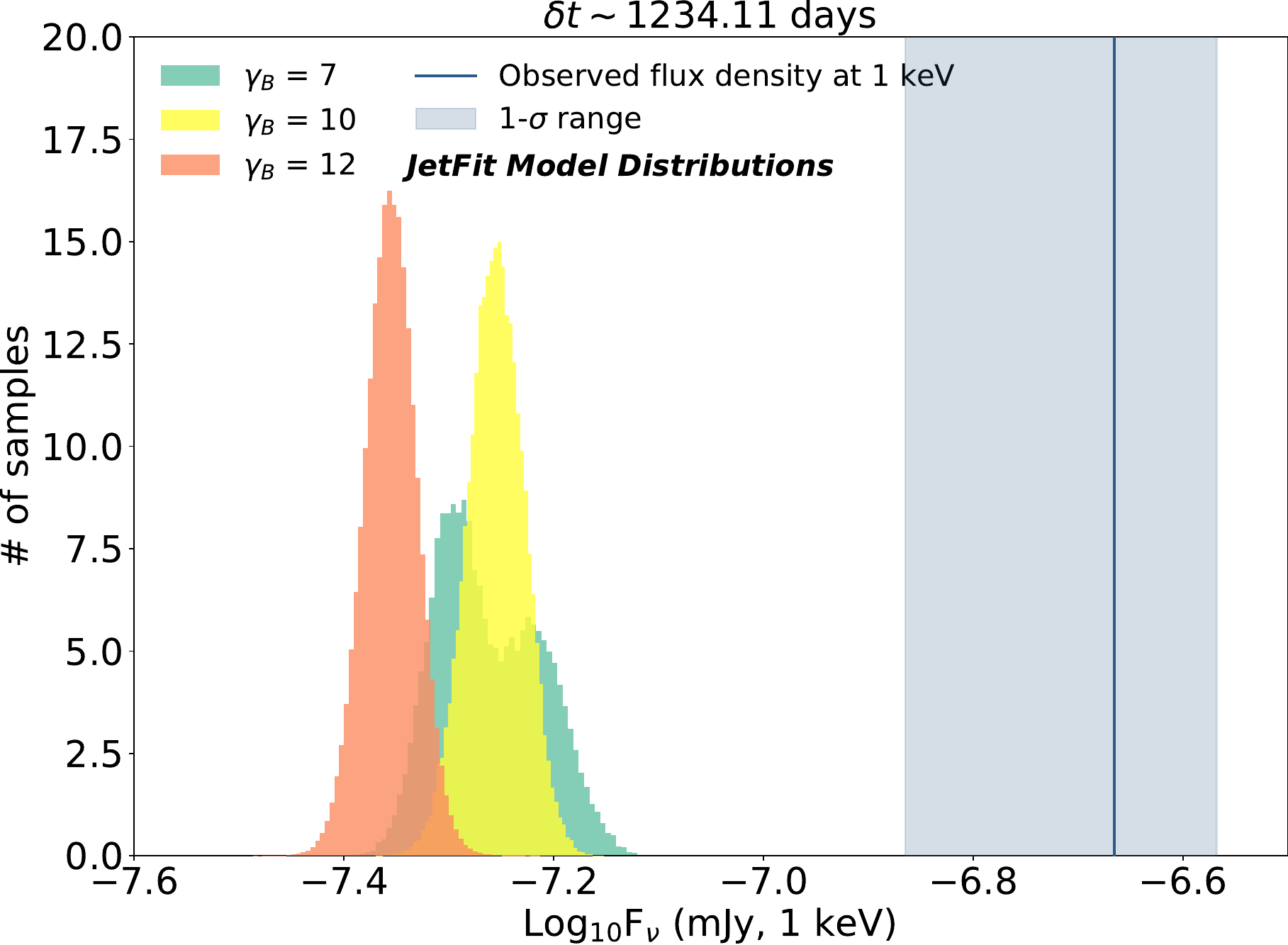}
\end{minipage}
\vspace*{5mm}%
\caption{
\emph{Top Panel:} Universal post jet-break model distributions. Expected 1-keV flux density distributions at 939.31 and 1234.11 days (histograms in color)  derived from fitting the post-peak multi-wavelength afterglow of GW170817 in the post jet-break regime with $F_{\rm \nu}\propto \nu^{-\beta}t^{-\alpha}$ (where $\beta=(p-1)/2$ and $\alpha=p$) in the time range $t_{\rm start}<\delta t< 900$ days for a variety of choices of $t_{\rm start}$. \emph{Bottom Panel: } \texttt{JetFit} model distributions. Expected 1-keV flux density distributions at 939.31 and 1234.11 days (histograms in color) derived from the fitting of the multi-wavelength afterglow of GW170817 in the time range $2 < \delta t < 900$ days using the code \texttt{JetFit} (using different values of $\gamma_{\rm B}$). \emph{Both Panels: }Vertical blue thick line and shaded area: observed X-ray flux density at the corresponding epoch and $\pm\,1\sigma$ confidence range, {for which the flux calibration was performed by conservatively assuming a jet-afterglow spectrum (\S\ref{SubSec:CXOspecanalysis}- \S\ref{Sec:Excess_stats}).}}
\label{Fig:postjetbreakstat}
\end{figure*}

{In this section, we calculate the significance of a deviation, \emph{if any}, of the latest X-ray observations from jet afterglow models. We perform our statistical analysis in the count phase-space to fully account for the Poissonian nature of the process.  Our results and conclusions do not depend on the specific flux calibration of the latest epochs of data at $> 900$\,days that we present in Table \ref{Tab:xrayanalysistab} with the purpose of offering to the reader a flux scale, and we explain the reasons as follows. Our goal is to test for the presence of a departure from the expectations of jet afterglow models, which have  solid predictions both in terms of temporal behavior and in terms of spectral behavior of the data at $\delta t>900$ days. Jet models can be violated spectrally, temporally, or both spectrally and temporally. The X-ray data at $\delta t>900$ days provide no useful spectral constraint (\S\ref{SubSec:CXOspecanalysis}). At this point we face two options: (i) we can perform a flux calibration of the X-ray count-rates at $\delta t>900$ days assuming a spectral model that is \emph{not} consistent with jet afterglows. This clearly violates jet afterglow models and implies that these models can be rejected at late times, and that the late-time X-rays come from a different source of emission; or (ii) we conservatively assume that the late-time X-rays have a spectrum that is consistent with jet afterglow models and we use the range of spectral models that are statistically allowed by jet afterglows to convert the predicted fluxes into observed count-rates on the CXO detector. We adopt approach (ii). This approach is conservative, in the sense that we are \emph{assuming} that the late-time X-ray spectrum is exactly as expected based on jet afterglow models, while in fact this might not be true. With this assumption, we then address the question: what is the probability of detecting a number of X-ray photons at least as large as the one observed at $\delta t>900$ days as a result of a statistical fluctuation of the model and background? }

{In the following we first adopt a jet-structure model agnostic approach, and use the universal post jet-break model to assess the potential deviation of the late-time X-ray data (\S\ref{SubSec:statUniversalJet}). Second, we use the off-axis structured jet afterglow models as computed by \texttt{JetFit} (\S\ref{SubSec:statJetFit}). Finally, we address in detail potential sources of systematic uncertainties and the performance of other numerical afterglow models. For the remainder of this section we note that including or not including in the initial jet afterglow fitting procedure the X-ray data at $\delta t>900$ days leads to differences that are smaller than the quoted level of precision of the statistical significances of the departure of the data from the models. This is a direct consequence of the limited number of X-ray photons at $\delta t>900$ days, which leaves the X-ray flux fundamentally  unconstrained (\S\ref{SubSec:CXOspecanalysis}). We elected to fit the data at $\delta t<900$ days, which has the advantage of preserving the statistical independence of the data acquired at $\delta t>900$ days from the models we are testing. We end by noting that from a statistical perspective we are not comparing two sets of models. Instead, we are assessing the potential departure of a sub-sample of data from models of jet afterglow emission. }

\subsection{Jet afterglows from the universal post jet-break model} \label{SubSec:statUniversalJet}
We first assessed the statistical evidence of an excess of X-ray emission with respect to the off-axis jet afterglow model by fitting the post-peak multi-wavelength afterglow decay with the following model $F_{\rm \nu}\propto \nu^{-\beta}t^{-\alpha}$. The X-ray to radio emission of GW\,170817 is powered by synchrotron radiation in the optically thin regime \citep{Margutti+2018binary, Fong+2019} for which $\beta=(p-1)/2$. 
Standard closure relations (e.g., \citealt{Lamb+2018}) in the post jet break phase, which apply to the post-peak afterglow evolution,  imply $\alpha = p$ \citep{Sari+1999}.  Our ``universal post jet-break model''  is thus: $F_{\rm \nu}\propto \nu^{-(p-1)/2}t^{-p}$. The multi-wavelength jet afterglow of GW\,170817 peaked at $t_{\rm pk}\approx 160$ days \citep{Dobie+2018turnover, Alexander+2017electromagnetic, Mooley+2018strong}. 

We fitted the multi-wavelength post-peak jet afterglow evolution with the model above in the time range $t_{\rm start}< \delta t<900$ days for several choices of start time $t_{\rm start}=157,163,172,196,209,215, 230$ days. 
We selected a range of $t_{\rm start}$ times starting from $t_{pk}$  to account for the unknown onset time of the asymptotic post-peak post jet-break power-law decay. We used VLA observations at 3 and 6 GHz compiled from \citealt{Hallinan+2017radiocounterpart, Alexander+2017electromagnetic, Mooley+2018mildly, Margutti+2018binary, Dobie+2018turnover, Alexander+2018, Mooley+2018superluminal, Hajela+2019}; \emph{Hubble Space Telescope} (\emph{HST}) observations at optical wavelengths from \citealt{Fong+2019}; and \emph{CXO} observations at 1 keV from this work. As an example, we show the plot of the best-fitting model, and the corresponding 68\% confidence interval, obtained assuming $t_{\rm start} = 196$\,days {in Figure \ref{Fig:JetfitLCcorner}. We assess the statistical significance of the departure of the late-time X-ray data for each choice of $t_{\rm start}$. }

We employed MCMC sampling with a Python module, \texttt{emcee} \citep{emcee2013}. For each choice of $t_{\rm start}$ we sampled $10^5$ times the expected X-ray flux density distribution at 1 keV ($F_{\rm 1\,keV}$) at the times of the last two \emph{CXO} epochs at  $t_1=939.31$ days and $t_2=1234.11$\,days ( Table\ref{Tab:xrayanalysistab}, top panel in  Figure \ref{Fig:postjetbreakstat}). For each MCMC sample we converted the predicted 1-keV flux densities ($F_{\rm 1\,keV,1}\equiv F_{\rm 1\,keV}(t_1)$ and $F_{\rm 1\,keV,2}\equiv F_{\rm 1\,keV}(t_2)$) into observed $0.5$ -- $8$\,keV total (i.e. source plus background) counts in a 1\arcsec\, region ($c_1$ and $c_2$) using the respective exposure times, the count-to-flux conversion factors derived from \emph{Xspec} and the observed background. We computed for each MCMC sample $i$ the probabilities $P_{\rm i,1}\equiv \rm{Pois}(c\ge N_{\rm obs,1}|c_1)$ and $P_{\rm i,2}\equiv \rm{Pois}(c\ge N_{\rm obs,2}|c_2)$, which represent the probability  of each sample to produce a number of X-ray photons larger or equal to those observed at $t_1$ and $t_2$ after \emph{Xspec} filtering in the $0.5$ -- $8$ keV energy band ($N_{\rm obs,1}=6$ and $N_{\rm obs,2}=12$, as noted in \S\ref{SubSec:CXOspecanalysis}) as a result of a Poissonian fluctuation. For each model defined by the choice of $t_{start}$, the total probability to lead to a deviation at least as prominent as the one observed at $t_1$ and $t_2$ is the re-normalized sum of the sample probabilities: $P_1= \frac{1}{N_{\rm sample}}\sum_i P_{\rm i,1}$ and $P_2=\frac{1}{N_{\rm sample}}\sum_i P_{\rm i,2}$. We find that the resulting $P_1$ and $P_2$ vary in the range $P_1=0.060$ -- $0.139$  and $P_2=2.61\times 10^{-4}$ -- $1.53\times 10^{-3}$  depending on the choice of $t_{\rm start}$. The observed X-rays at 1234 days thus correspond to a $3.2\,\sigma$ -- $3.7\,\sigma$ (Gaussian equivalent, 99.8626\% - 99.9784\%)\footnote{Probabilities in the form of \% added as per referee\#2 request.} deviation from the off-axis jet model. $P_1\times P_2$ thus lies in the range $P_1\times P_2=1.73\times 10^{-5}$ -- $2.50\times 10^{-4}$, {where the range of probabilities reported reflect the assumed $t_{\rm start}$ {(Table  \ref{tab:Pchance})}.   Finally, the combined probability to obtain deviations from the {universal post-jet break} off-axis  model 
at 939.31 days and 1234.11 days can be conservatively estimated as $P_{\rm combined}\equiv \frac{1}{N_{\rm{sample}}} \sum_i \rm{Pois}(c\ge (N_{\rm obs,2}+N_{\rm obs,1})|(c_{1,i}+c_{2,i}))$. We find $P_{\rm combined}=9.51\times 10^{-5}-1.37\times 10^{-3}$ (3.2-3.9$\sigma$, Gaussian equivalent, depending on the choice of $\mathbf{t}_{\rm \mathbf{start}}$).
We conclude that the observed X-rays at $\delta t>900$ days represent a statistical deviation from the expectations of the universal post jet-break models that best fit earlier observations of GW\,170817 with statistical significance $\ge 3.2\sigma$. The chance probabilities as a function of $t_{\rm start}$ are reported in Table \ref{tab:Pchance}.}

\subsection{Jet afterglows computed with \texttt{JetFit}} \label{SubSec:statJetFit}

\begin{figure*}[!t]
\begin{center}
\hspace*{0.1in}
\hspace*{-0.8in}
\includegraphics[scale=0.28]{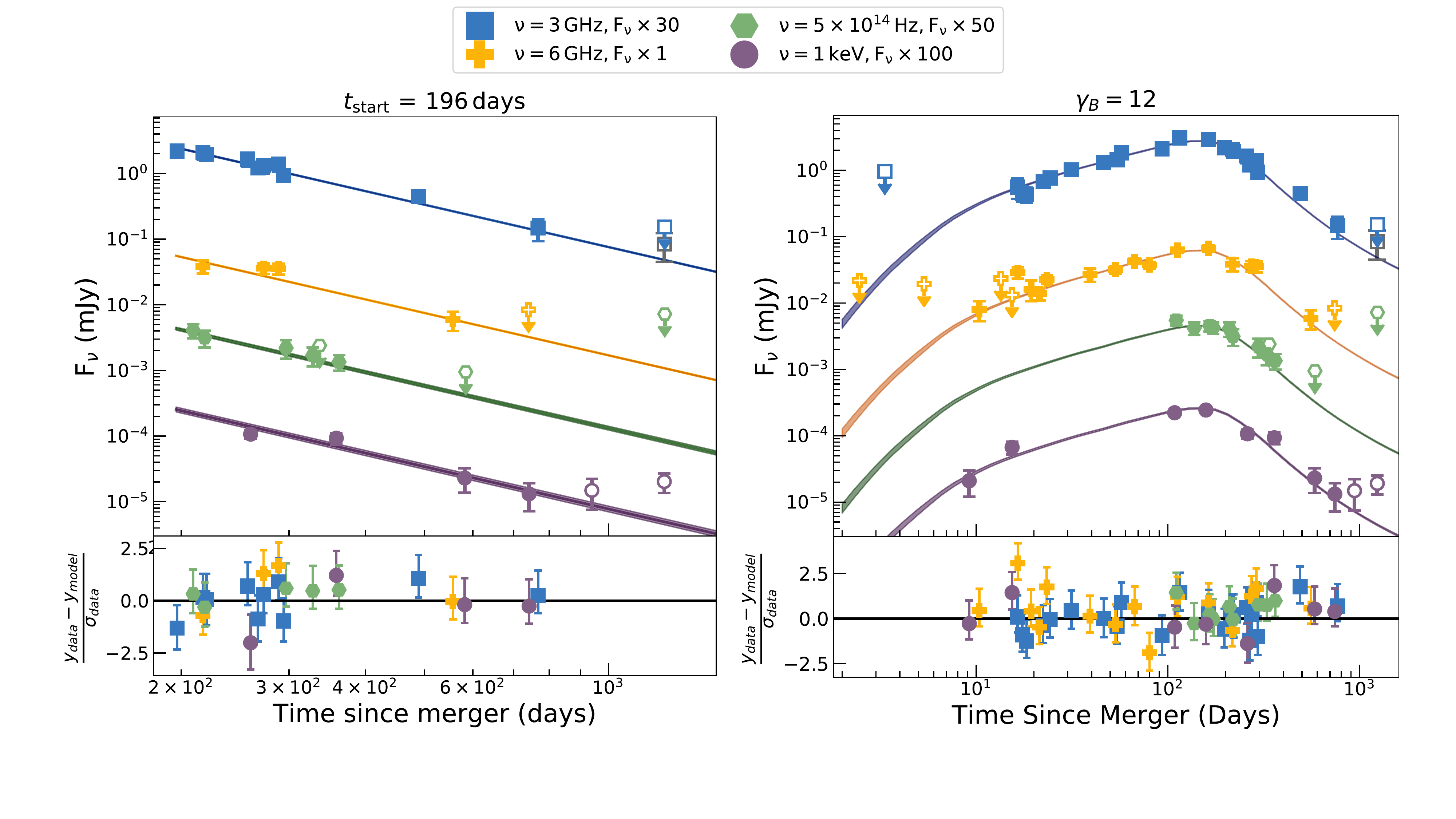}
\vspace{-0.5in}
\caption{\emph{Left Panel:} {Non-thermal emission from GW\,170817 across the electromagnetic spectrum and best fitting universal post jet-break model, as explained in \S\ref{Sec:Excess_stats},  with $t_{\rm start} = 196$\,days. 
\emph{Right Panel:} Non-thermal emission from GW\,170817 and best fitting jet-afterglow model computed with \texttt{JetFit} for $n=0.01\,\rm{cm^{-3}}$, $\epsilon_{\rm e} = 0.1$, and $\gamma_{\rm B} = 12$ fixed. In both panels empty symbols were not included in the fitting procedure but are shown here for completeness adopting a flux calibration consistent with the afterglow models. Colored bands identify the 68\% flux confidence interval.}
The grey empty square symbol is the peak pixel value within one synthesized beam at the location of GW\,170817 at 3 GHz from \citealt{Balasubramanian21}. {\textcolor{black}{The bottom subplots show the the difference between observations and the best-fitting models as derived from the model posteriors, and expressed in units of $1\sigma$ data uncertainties for displaying purposes.  
}}}

\label{Fig:JetfitLCcorner}
\end{center}
\end{figure*}

We further performed a similar statistical study to test the excess of X-ray emission with respect to the off-axis structured jet light-curves modeled with \texttt{JetFit} \citep{Duffell+2013, Ryan+2015, Wu+2018}. \texttt{JetFit}  fits the afterglow light curves for arbitrary viewing angles using a `boosted-fireball' structured jet model to compute the jet dynamics as it spreads. It naturally accommodates a diverse range of outflows from mildly-relativistic quasi-spherical outflows to ultra-relativistic highly collimated jets. \texttt{JetFit} uses the python package \texttt{emcee} to explore the full parameter space formed by eight parameters: the explosion energy, $E_0$; the ambient density, $n$; the asymptotic Lorentz factor, $\eta_0$; the boost Lorentz factor, $\gamma_{\rm B}$; the spectral index of the electron distribution, $p$; the electron energy fraction, $\epsilon_{\rm e}$; the magnetic energy fraction, $\epsilon_{\rm B}$; and viewing angle $\theta_{\rm obs}$; and finds the best-fitting values and their posterior distributions.
  Because the broadband SED (spectral energy distribution) of GW\,170817, from $\delta t = 2 - 745$ days, is best explained by a simple power-law, some of these parameters are highly degenerate and the problem is under-constrained. Hence, we fixed $\epsilon_{\rm e} = 0.1$, as predicted from the simulations of particle acceleration by relativistic shocks \citep{Sironi+2013maximumenergy}, $n = 0.01\,\rm{cm^{-3}}$, the upper-limit on the ambient density inferred from the study of the host X-ray thermal emission \citep{Hajela+2019}, and {we computed the best-fitting models assuming three values of} $\gamma_{\rm B} = 7,10$, and $12$. 
We selected these $\gamma_{\rm B}$ values based on the VLBI measurements of the angular displacement of the radio emission with time, which  constrained the jet Lorentz factor $\Gamma \approx 4$ at the time of the afterglow peak (or $\theta_{\rm obs}-\theta_{\rm j}\approx 1/\Gamma\approx 1/4$,  \citealt{Mooley+2018superluminal}).\footnote{\texttt{JetFit} can reliably predict the afterglow from boosted fireballs with $\gamma_{\rm B}\le 12$, which translates into  $\theta_{\rm j}\approx 1/\gamma_{\rm B}\ge 4.8^\circ$}
We use \texttt{JetFit} to fit the multi-wavelength afterglow light-curves at  $3$\,GHz, $6$\,GHz, optical  and at $1$ keV frequencies
acquired at $2 < \delta t < 900$ days.
The jet opening angle $\theta_{\rm j}$ of GW\,170817 has been estimated to be of the order of a few degrees \citep{Mooley+2018superluminal,Ghirlanda+2019,Nathanail+2021}, and we thus consider the $\gamma_{\rm B}=12$ boosted fireball model as our fiducial case.  The best-fitting light curves for the  {$\gamma_{\rm B} =12$ are shown in  Figure \ref{Fig:JetfitLCcorner}, while  the one- and two-dimensional projections of the posterior distribution of the free parameters for $\gamma_{\rm B} =12$ are provided in Appendix \ref{Appendix2}, Figure \ref{Fig:postjetbreakmodelfit}}.

We use the full posterior distribution of all the free parameters to compute the distribution of flux density at 1 keV at $t_1$ and $t_2$ {for each choice of $\gamma_{\rm B}$. } Similar to the above statistical analysis, we convert these flux densities to the total counts in the $0.5$ -- $8$ keV energy range in a 1\arcsec\, region, calculate the probability of each sample, i, $P_{\rm i,j} = \rm{Pois}(c \ge N_{\rm obs,j}|c_{\rm j})$, where $\rm{j} \in {1,2}$ for the two epochs respectively, and finally compute the cumulative probabilities, $P_{\rm j}$, to lead to a deviation at least as prominent as the one observed at $t_{\rm j}$ (bottom panel,  Figure \ref{Fig:postjetbreakstat}). For different values of $\gamma_{\rm B}$, we find $P_{\rm j}$ in the range $P_1 = 0.07$ -- $0.15$ and $P_2 = 7.36 \times 10^{-4}$ -- $2.82 \times 10^{-3} $, corresponding to a $2.9\,\sigma$ -- $3.4\,\sigma$ (Gaussian equivalent, 99.6268\% - 99.9326\%) deviation of the observed X-rays at 1234 days from the light-curve modeled by the off-axis structured jet model. We further find 
$P_1\times P_2= 5.59 \times 10^{-5}$ -- $4.69 \times 10^{-4}$  ($3.5\,\sigma$ -- $4.0\,\sigma$ Gaussian equivalent, 99.9535\% - 99.9937\%), and $P_{\rm{combined}}=2.62 \times 10^{-4}$ -- $2.10 \times 10^{-3}$ {($3.1\,\sigma$ -- $3.7\,\sigma$, where the range reflects the assumed values of $\mathbf{\gamma}_{\rm \mathbf{B}}$ used)} to obtain deviations from the off-axis structured jet model at least as prominent as those observed at both epochs  $t_1$ and $t_2$. Larger $\gamma_{\rm B}$ values imply a higher level of collimation of the jet, and hence a faster post-peak transition to the asymptotic power-law decay, which explains the highest significance of the excess associated to the $\gamma_{\rm B}=12$ model (bottom panel,  Figure \ref{Fig:postjetbreakstat}). Since for GW\,170817 $\theta_j\le5^\circ$ (e.g., \citealt{Mooley+2018superluminal, Ghirlanda+2019}) and our most collimated model has $\gamma_{\rm B}=12$ (i.e. $\theta_j\approx 5^\circ$), in this sense the probabilities derived with this approach are conservative.  {For the same reason, $\gamma_{\rm B}=12$ is our baseline model and for  this set of models the probabilities associated with $\gamma_{\rm B}=12$  should be considered the most realistic estimates (i.e. $P$ of chance deviation corresponding to $3.7\sigma$).  The chance probabilities as a function of $\gamma_{\rm B}$ are reported in Table \ref{tab:Pchance}.}

{From Figure, right panel, \ref{Fig:postjetbreakstat}, the $\gamma_{\rm B}=12$ best fitting model lies slightly above the central value of the data points at $\delta t>300$ days, but it is well within the $1\sigma$ error bars of the data at $\delta t >600$ days and always within the $2\sigma$ range. For this model the $\chi^2/dof=1.03$. We further tested the departure from a random distribution of the signs of the residuals implementing a Runs test for randomness. Our data set contains 54 data points and the number of runs is 24. The  chance probability of obtaining the observed distribution of runs is $26$\%. It follows that the hypothesis of random distribution of the model's residuals cannot be rejected.     }

\subsection{General Considerations} \label{SubSec:statGenCons}
Both statistical approaches detailed above (i.e. the jet afterglow light-curve models and the universal post jet-break power-law decay) independently lead to the conclusion of the presence of an X-ray excess of emission at $\delta t>900$ days with statistical confidence $\ge 3.1\,\sigma$.  Observations acquired around $940$ days alone do not provide any statistically significant evidence of a deviation from the expectations of an off-axis jet model as we reported in \cite{Hajela+2020} (see also \citealt{Troja+2020}).
The statistical significance of the excess of X-ray emission is driven by our most recent epoch of \emph{CXO} data at $1234$ days. 
We also note that we do \emph{not} claim a re-brightening of the X-ray flux, but a statistically significant deviation from the existing models that best fit the afterglow at $<900$\,days, which points to the emergence of a new X-ray component. {Our approach is agnostic with regard to the spectral and temporal properties of any additional emission component. The statistical tests that we carried out have been explicitly designed to avoid any dependency on any assumed property of the additional component and instead test for a deviation compared to expectations from the jet afterglow emission.}

We note that systematic uncertainties on the relative flux calibration of \emph{Chandra}/ACIS-S between observations acquired at $\delta t<900$ days and $\delta t>$900 days have minimal impact on our conclusions. We use the \texttt{JetFit} models with  $\gamma_B = 12$ here as an example to quantify this effect. Specifically, adopting a systematic RMS flux variation of $<3.4\%$ (Chandra calibration team, private communication) on \emph{Chandra}/ACIS-S fluxes, and assuming that fluxes at $\delta t> 900$ days have been systematically overestimated by that RMS factor, we find evidence for a $3.97\,\sigma$ (Gaussian equivalent) deviation of the X-rays emission at $\delta t > 900$\,days from the best fitting model. Similar results hold for the universal post-jet break afterglow models. Finally, we note that correcting or not correcting for PSF-losses the count-rates at $\delta t>900$ days induces changes in the stated statistical significances of the excess of $\lesssim 0.1\,\sigma$ (Gaussian equivalent), demonstrating that the choice of the aperture size of $1$\arcsec\, does not drive our conclusions.

We end by addressing the difference between our conclusions and the claim of a statistical significance of the X-ray excess of $\lesssim 3\,\rm{\sigma}$ that appeared in \cite{Troja+2021}. 
The main source of difference between our analysis and the one presented in \cite{Troja+2021}, which drives the different conclusions about the statistical evidence for an X-ray excess, is related to the statistical treatment of the data, to the Poisson nature of the X-ray signal at $t> 900$ days, and to the specific jet model chosen as a reference. Specifically, the use of a jet model presented in their work that is \emph{not} in tension with the VLBI measurements would have led to the inference of a significantly larger discrepancy between X-ray observations at $\delta t>900$ days and expectations as we demonstrate in Figure \ref{Fig:xraycomparison} in Appendix \ref{Appendix}. The X-ray flux calibration plays a negligible statistical role, as we show in Figure \ref{Fig:xraycomparison} and Appendix \ref{Appendix}. Our statistical tests self-consistently account for the Poisson nature of the process, using jet models that are \emph{not} in violation of the VLBI constraints.
{To the extent of the authors' knowledge, there is no jet model that does not violate the VLBI constraints and can naturally reproduce the late time X-rays of GW\,170817.}

\section{Inferences on the Broadband Spectrum at 1234 days}
\label{Sec:SpecEvol}
\begin{figure}[!t]
\begin{center}
\scalebox{1.}
{\includegraphics[width=0.5\textwidth]{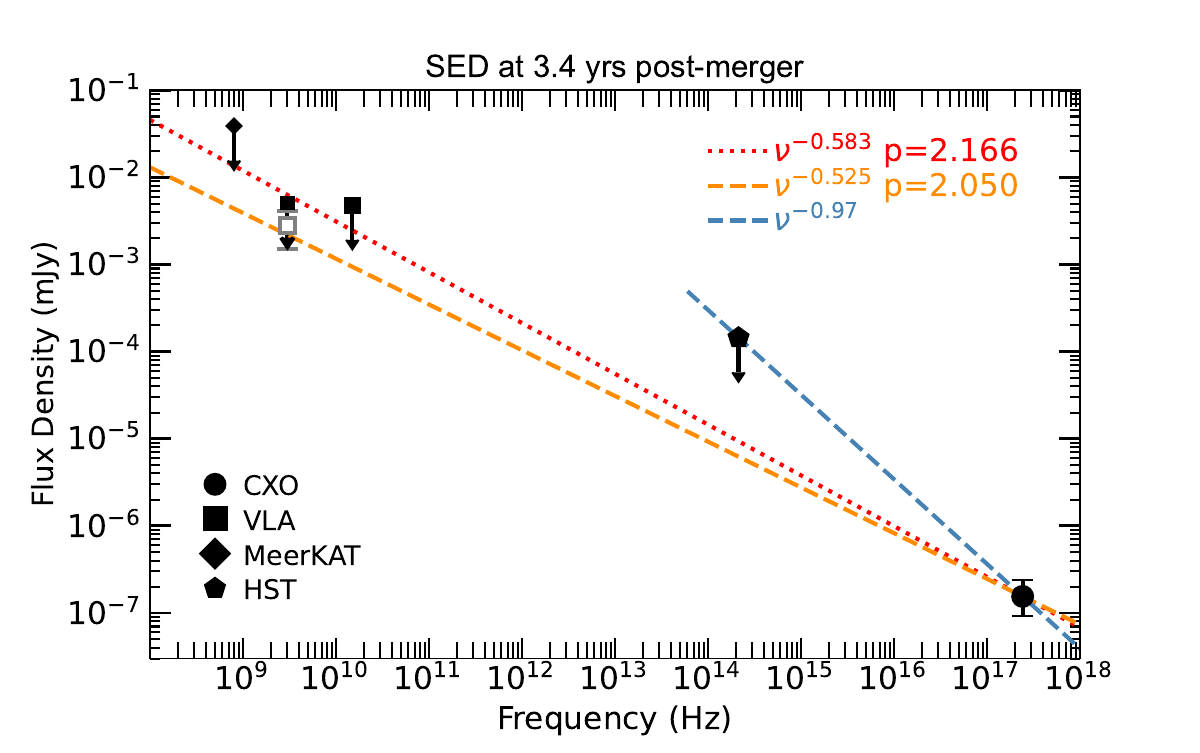}}
\caption{Broad-band spectral energy distribution acquired around $\delta t\approx$3.4 years post-merger, including CXO X-ray data (filled circle), VLA upper limits at 3 and 15 GHz (filled squares),  MeerKAT flux limit (filled diamond) and \emph{HST}/F140W flux limit (filled hexagon). Grey open square: 3 GHz peak flux pixel value of $2.8\,\rm{\mu Jy}$ (with RMS of $1.3\,\rm{\mu Jy}$)  within one synthesized beam at the location of GW\,170817 from \citealt{Balasubramanian21}.
Red dotted line: $F_{\rm \nu}\propto \nu^{-(p-1)/2}$ spectrum with $p=2.166$ that best fitted the jet-afterglow data \citep{Fong+2019}. The VLA 3 GHz limit suggests a shallower spectrum (\S\ref{Sec:SpecEvol}). Orange dashed line: $F_{\rm \nu}\propto \nu^{-(p-1)/2}$ with $p=2.05$. \emph{HST} observations imply a NIR-to-X-ray spectral slope steeper than $\approx 1$.}
\label{Fig:SED}
\end{center}
\vspace{-0.5cm}
\end{figure}

\begin{figure}[!t]
\begin{center}
\scalebox{1.}
{\includegraphics[width=0.5\textwidth]{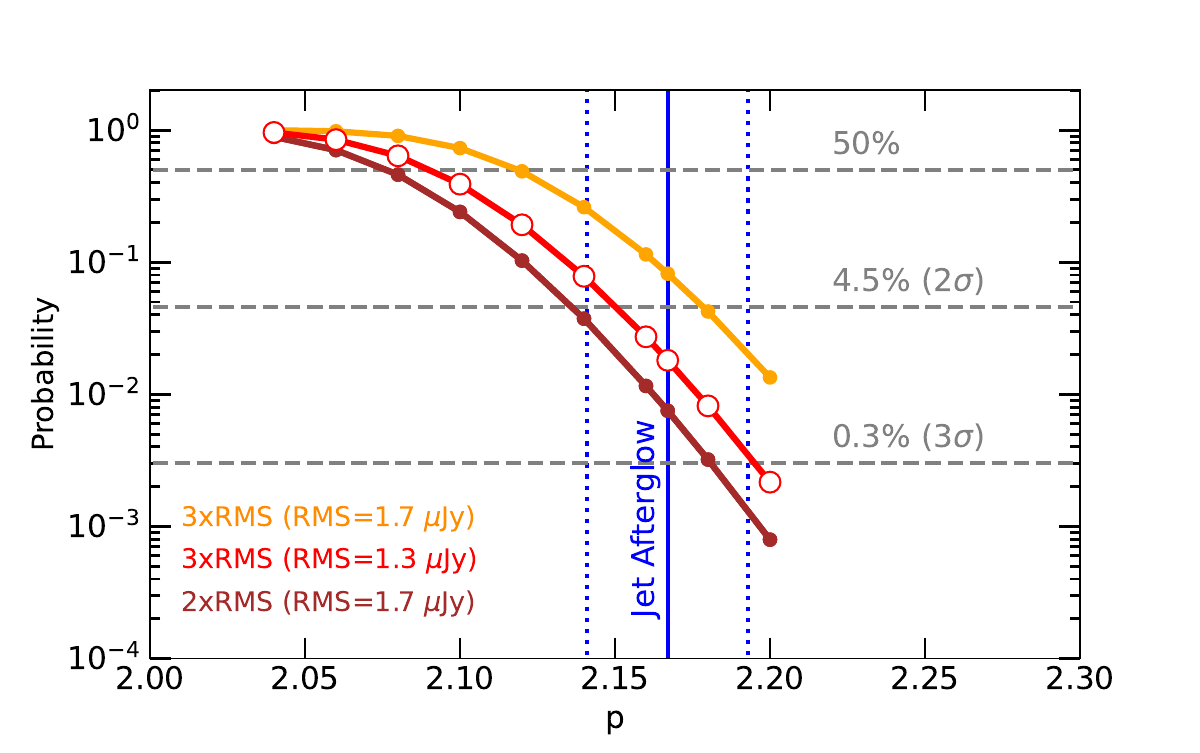}}
\caption{Probability of simple power-law $F_{\rm \nu}=\rm{\mathbf{Norm}}\times\nu^{-(p-1)/2}$ spectral models at 1234 days that do not violate the  $3\times$RMS (orange), and $2\times$RMS (brown) flux density of our 3\,GHz image at the location of GW\,170817 as a function of $p$, where RMS=$1.7\,\mu Jy$ inferred from this work. Red line and open symbols: results for RMS=$1.3\,\mu Jy$ inferred by \citealt{Balasubramanian21}, $\rm{\mathbf{Norm}}$ is drawn from the posterior probability distribution of the $0.3$ -- $10$ keV unabsorbed X-ray flux at 1234 days as derived from MCMC sampling within \emph{Xspec}. Horizontal grey dashed lines mark the $0.3\%$, $4.5\%$ and $50\%$ probability levels. Vertical blue thick and dotted lines: best fitting $p$ parameter and $1\,\sigma$ range for the jet afterglow  as derived from broad-band SED fitting of the non-thermal emission of GW\,170817 at $\delta t<900$ days \citep{Fong+2019}. This analysis suggests a hardening of the non-thermal spectrum of GW\,170817 at 1234 days to values of $p<$ than the best-fitting value from the earlier jet afterglow at statistical confidence $\ge 92\%-99.2\%$.} 
\label{Fig:SpecEvol1234days}
\end{center}
\vspace{-0.5cm}
\end{figure}

The broadband X-ray-to-radio non-thermal emission from the jet afterglow of GW\,170817 at $\delta t<900$ days is well fitted by a simple power-law spectral model $F_{\rm \nu}\propto \nu^{-\beta}$ with $\beta = 0.583 \pm 0.013$ \citep{Fong+2019}, or equivalently, $F_{\rm \nu}\propto \nu^{-(p-1)/2}$ with $p = 2.166 \pm 0.026$ in the optically thin synchrotron regime. In this section we compute the constraints on the spectral slope at $= 1234$\,days that are imposed by the X-ray detection (\S\ref{Sec:xray})   and the 3\,GHz radio limits (\S\ref{SubSec:VLA_obs}) under the assumption that the broadband spectrum is still described by a simple power-law model. Radio limits at 15 GHz and 0.8 GHz (\S\ref{SubSec:VLA_obs}-\ref{SubSec:MeerKAT_obs}), and \emph{HST} observations \citep{KilpatrickGCN+2021}\, do not provide additional constraints on the simple power-law model ( Figure \ref{Fig:SED}).  
 We used MCMC sampling within \emph{Xspec} as described in \S\ref{SubSec:CXOspecanalysis} and we sampled $10^6$ times the posterior probability distribution of the unabsorbed $0.3$ -- $10$\,keV flux derived from fitting the \emph{CXO} data at $ 1234$\,days employing Cash-statistics. This method accounts for deviations from Gaussian statistics that manifest in the regime of low spectral counts. We then computed as a function of $p$ the probability associated with spectral models $F_{\rm{\nu}} \propto \nu^{-(p-1)/2}$ that would not lead to a radio detection, here defined as a 3\,GHz radio flux density above $\mathbf{3\times}$,  or alternatively $\mathbf{2\times}$, the flux density root mean square -- RMS --  of our image around the location of GW\,170817, where RMS $= 1.7\,\mu Jy$. 

Our results are shown in  Figure \ref{Fig:SpecEvol1234days}. We find that values of $p>$ 2.166, i.e. larger than the best fitting value of the jet-afterglow at $\delta t<900$ days are ruled out with statistical confidence $\ge 92\%-99.2\%$. These results suggest the evolution of the broadband spectrum towards lower values of $p$ and constitute the first indication of spectral evolution of the non-thermal emission from GW\,170817. This conclusion is strengthened by using the RMS $= 1.3\,\rm{\mu Jy}$ at 3\,GHz from \citealt{Balasubramanian21}.
We end by noting that \emph{HST} observations acquired on $\delta t=$1236.5 days since merger at $\nu=2.13\times 10^{14}\,\rm{Hz}$ \citep{KilpatrickGCN+2021} imply an optical to X-ray spectral index $\beta_{\rm OX}\lesssim 0.97$ (where $F_{\rm \nu}\propto \nu^{-\beta_{\rm OX}}$). Finally, our VLA observations at 15\,GHz reach a similar depth as our 3\,GHz  observations and rule out an optically thick $F_{\nu}\propto \nu^2$ radio source with flux density $F_{\rm \nu}\ge \rm{0.06\,\mu Jy}$ at $\nu = 3$\,GHz.

\section{Late time evolution of the emission from off-axis jet afterglows}
\label{Sec:latetimejet}
In the context of synchrotron emission from an ultra-relativistic off-axis jet, a post-peak late-time flattening  of the light-curve can be the result of: (i) the jet encounter with an over-density in the environment; (ii) energy injection; (iii) time-varying shock microphysical parameters $\epsilon_{\rm B}$ and $\epsilon_{\rm e}$; (iv) transition into the sub-relativistic phase; and (v) emergence of the counter-jet emission \citep{Granot+2018,NakarPiran2021}.

The universal post jet-break light-curve evolution for an observed frequency $\nu$ above the synchrotron self-absorption frequency $\nu_{\rm sa}$ and for  $\nu_m<\nu<\nu_c$ (where $\nu_{\rm m}$ is the synchrotron frequency and $\nu_c$ is the cooling frequency) is \citep{Granot+2018}: 
\begin{eqnarray}
F_{\rm \nu}(\nu,t)\propto \epsilon_{\rm e}^{p-1}\epsilon_{\rm B}^{\frac{p+1}{4}} n^{\frac{3-p}{12}}E_k^{\frac{p+3}{3}}t^{-p}\nu^{\frac{1-p}{2}}
\label{Eq:postpeakflux}
\end{eqnarray}
where $E_k$ is the jet energy and $n$ is the circum-burst density.
The observed X-ray emission at 1234 days is a factor $\approx 4$ above the extrapolation of the off-axis jet afterglow models (Figure \ref{Fig:Xrays}). Explaining this excess of emission as a result of an over-density in the environment would require an exceedingly steep density gradient with $n$ increasing by a factor of $ (4)^{\frac{12}{3-p}}\approx 3\times 10^8$ (Eq. \ref{Eq:postpeakflux}) over $\Delta r/r\approx 1$ at $r\approx1$ pc. 
The characteristic size of the bow-shock cavity inflated by a pulsar wind (\emph{if} any of the NS progenitors of GW\,170817 was a pulsar) scales as $R_s\propto n_{\rm ext}^{-1/2}$, where $n_{\rm ext}$ external medium density probed by the wind \citep{Ramirez+2019}. Following \citet{Ramirez+2019}, their equation 4, $R_s$  is expected to be  a factor $\gtrsim 3-8$ smaller than the
the shock radius at this time if the density probed by the jet $n=10^{-4}-10^{-2}\,\rm{cm^{-3}}$ is representative of the density in the evacuated region (as it is reasonable to expect $n_{\rm ext}>n$). 
Additionally, for a density contrast $\approx 10^{8}$ the implied amount of mass at $r\approx 1$ pc within the jet angle is $\ge 10\,\rm{M_{\odot}}$. We thus consider the jet encounter with the edge of an associated pulsar wind bubble unlikely to occur at the time of our monitoring.
Deep \emph{HST} observations of the host galaxy environment of GW\,170817 rule out the presence of a globular cluster (GC) at the location of BNS merger \citep{Fong+2019,Blanchard+2017electromagnetic,Levan+2017,Pan+2017,Lamb+2019}.  
The gravitational potential well of a GC might otherwise provide a physical reason for an abrupt change in the external gas density on the scale probed by the afterglow.  We thus do not consider the over-density scenario any further. 

Following a similar line of reasoning, an excess of emission can be produced if the shock is re-freshed by the deposition of new energy (e.g. \citealt{Laskar+2015,SariMeszaros2000}).
 From Eq. \ref{Eq:postpeakflux}, a flux ratio of $\approx 4$ requires the late-time deposition of a large amount of additional energy similar to the jet energy $E_k$. There is no plausible  energy source that can power the sudden energy release of an amount of energy equivalent to the jet energy at late times and we consider this scenario unlikely.  Finally, a sharp variation of the shock microphysical parameters $\epsilon_{\rm e}$ and $\epsilon_{\rm B}$ with time can in principle lead to larger fluxes. This scenario would require an ad hoc evolution of $\epsilon_{\rm e}$ and $\epsilon_{\rm B}$ to explain the X-ray observations and we thus consider this model not physically motivated. Additionally, the deceleration of the shock is expected to lead to smaller $\epsilon_e$ values, while larger $\epsilon_e$ values would be needed to explain a flatter light-curve. In addition to the arguments above, we end by noting that all the models discussed so far do not naturally explain the harder radio-to-X-ray spectrum with a reduced value of $p$ (\S\ref{Sec:SpecEvol}).

In the absence of energy injection, environment over-densities and variations in the shock microphysical parameters, the transition of the blast wave dynamics to the sub-relativistic phase at 
$t_{\rm NR} \approx 1100 (E_{\rm k,iso,53}/n)^{1/3}$ days \citep{Piran2004} is expected to lead to a smooth transition to a less steeply decaying light-curve $F_{\rm \nu} \propto t^{-3(p-1)/2+3/5}$  at  $\nu_m<\nu<\nu_c$ (Equation 97, \citealt{Piran2004}) or $F_{\rm \nu} \propto t^{-3(p-1)/2+1/2}$ at  $\nu>\nu_c$ (equation A20, \citealt{Frail+2000}). For $p=2.05-2.15$ we expect the light-curve to decay as $F_{\rm \nu}\propto t^{-1.2}-t^{-1.0}$ in the non-relativistic regime.  For the jet-environment parameters of GW\,170817 (\citealt{Mooley+2018superluminal,Hotokezaka+2019,Ghirlanda+2019,Ryan+2020}
,  Figure \ref{Fig:JetfitLCcorner})  
the full transition to the non-relativistic regime and the appearance of the counter jet is expected  at  $t_{\rm NR}\ge5000$ days, significantly later than our current epoch of observation, with the start of the ``deep Newtonian phase'' being at even later times. 
 In the deep Newtonian phase $F_{\rm \nu}\propto t^{-3(1+p)/10}$ or $F_{\rm \nu}\propto t^{-0.9}$ for $p=2.05-2.15$ \citep{Sironi+2013latetimeflattening}.  A smooth transition to the sub-relativistic regime, accompanied by a slower light-curve decay,  might start to be noticeable at earlier epochs, and possibly now, as the jet-core bulk Lorentz factor is $\Gamma(t) \approx 4(t/100\,\rm{days})^{-3/8} \approx 1.6$ at the current epoch (still in the Blandford-McKee regime, no jet spreading) or    $\Gamma(t) \propto t^{-1/2}$ leading to $\Gamma(t)\approx 1.1$ for exponential jet spreading \citep{Rhoads1999}. These estimates are based on the inferred $\Gamma \approx 4$ at $\approx$ 100 days \citep{Mooley+2018superluminal}.   
In both cases the light-curve evolution is expected to be achromatic and the emission is expected to become dimmer with time as $F_{\rm \nu}\propto t^{-1}$ or steeper.  No excess can be explained within the non-relativistic jet transition scenario  and no spectral evolution is expected unless we invoke an ad hoc temporal evolution of $p$ from $p=2.15$ to $p=2.0$ in the time range $900-1200$ days (i.e. well before the full transition to the non-relativistic phase) as the shock decelerates. The theoretical predictions from the Fermi process of particle acceleration in shocks would support this trend of evolution, as they predict $p=2$  at non-relativistic shock speeds \citep{Bell1978,BlandfordOstriker1978,BlandfordEichler1987}\,  and $p\approx 2.22$\, at ultra-relativistic velocities in the test particle limit \citep{Kirk+2000,Achterberg+2001,Keshet+2005,Sironi+2013maximumenergy}.  However, here the challenge is represented by having a shock where the index of the non-thermal electron distribution $p$ changes with time as a result of the shock deceleration, without having a substantial drop in the electron acceleration efficiency $\epsilon_{\rm e}$ when compared to the earlier  ultra-relativistic regime \citep{Crumley+2019}. 
Finally, the emergence of the counter-jet emission is expected to lead to a flatter light-curve at $\delta t> t_{\rm NR}$, or $\delta t> 5000$ days for the parameters of GW\,170817.

To summarize, the late time evolution of the jet does not naturally account for the brightness, spectrum  and flattening of the X-ray light-curve at $\delta t \approx 1200$ days. Specifically: 
(i) the steep density gradient of a factor of $\approx 10^8$ over a pc scale required to explain the X-ray excess of emission implies an extremely large shell mass $\ge 10 M_{\odot}$ within the jet angle at $\approx 1$\,pc, making the scenario of a jet encounter with the edge of an associated pulsar-wind bubble unlikely;
(ii) similarly, the large amount of energy required to be injected to produce a X-ray excess is equivalent to the energy of the jet itself and there is no plausible source to power such an energy release at these late-times;
(iii) a sudden variation of the shock microphysical parameters is not physically motivated at this epoch; (iv) the shock transition to the Newtonian regime is expected to happen at significantly later times $t_{\rm{NR} \ge 5000}$\,days and no effect related to the Newtonian transition can thus be invoked to explain the late-time excess of X-ray emission; 
(v) lastly, the counter-jet is also expected to emerge at $\delta t > 5000$\,days.

\section{Kilonova Afterglow Models and Numerical Relativity Simulations of BNS Mergers}
\label{Sec:KNmodels}
NS merger simulations predict the ejection of neutron-rich and neutron-poor matter due to a variety of mechanisms operating over different timescales before, during and after the merger \citep{Shibata:2019wef}.  These mass outflows shock the circumbinary medium producing synchrotron radiation that peaks on the deceleration time scale $t_{\rm dec}$ \citep{Nakar+2011Nat}. The direct implication is that heavier mass outflows like those associated with the kilonova ejecta will produce non-thermal emission that will peak later in time than the emission associated with the significantly faster but also significantly lighter jet. For the inferred kilonova ejecta properties 
of GW\,170817 ($M_{\rm ej}\approx0.06\,\rm{M_{\odot}}$, $n\approx0.01-0.001\,\rm{cm^{-3}}$ and $\beta\approx 0.1-0.3$; \citealt{Villar+2017combined,Cowperthwaite+2017EM,Arcavi_2018,Waxman+2018,Nicholl+2021,Bulla+2019,Drout+2017LC,Kilpatrick+2017})),\, $t_{\rm dec}\approx 10^{4}$ days.
However, the deceleration of the fastest-moving tail of these ejecta is expected to contribute to non-thermal emission on significantly shorter timescales of months to years after the merger \citep{Nakar+2011Nat, Kyutoku+2014ultrarel, Takami+2014highene, Hotokezaka:2015eja, Hotokezaka:2018gmo,Kathirgamaraju+2019,MargalitPiran2020} that are relevant now  (while the bulk of slower-moving ejecta powered the UV/optical/IR kilonova at $\delta t<70$ days).  

This kilonova afterglow will appear as an excess of emission compared to the off-axis jet afterglow. Being powered by a different shock and by a different electron population than the jet's forward shock, the synchrotron emission from the kilonova afterglow does not necessarily inherit the same microphysical parameters $\epsilon_{\rm e}$, $\epsilon_{\rm B}$, as well as the electron index $p$. In this respect, the lower $p$ value indicated by our observations (Figure \ref{Fig:SpecEvol1234days}) would be a natural outcome and would be consistent with the $p<2.2$ theoretical expectation of shocks that are non-relativistic \citep{Bell1978,BlandfordOstriker1978,BlandfordEichler1987}.

The luminosity and time evolution of the kilonova afterglow from a BNS merger depends on (and is a tracer of) the intrinsic parameters that include how the ejecta energy is partitioned in the velocity space $E_{\rm{KN}}(\Gamma \beta)$, which ultimately depends on the neutron star equation of state (EoS) and the binary mass ratio $q$, and also the extrinsic parameters that include those that regulate the kilonova shock microphysics (fraction of post-shock energy density in relativistic electrons, $\epsilon_{\rm e,KN}$, and in magnetic field, $\epsilon_{\rm B,KN}$ and $p_{\rm{KN}}$), and the environment density $n$. 
We first adopt in \S\ref{SubSec:AdithanKN} an analytical parametrization of $E_{\rm{KN}}(\Gamma \beta)$ to explore the large parameter space of the kilonova afterglow parameters while being agnostic to the ejecta type (e.g. winds vs.  dynamical etc.). In the second part (\S\ref{SubSec:NRKN}) we employ a set of numerical relativity simulations of BNS mergers to emphasize the dependency of the observed kilonova afterglow on intrinsic parameters of the NS binary, like the binary mass ratio or the NS EoS. 
We note that the potential early emergence of the kilonova afterglow a few years after the merger, at a time when the jet has yet to effectively become spherical (\S\ref{Sec:latetimejet}) implies that the kilonova shock is expanding into a medium that is mostly unperturbed (i.e., not shocked by the jet shock) and that  effects related to the jet evacuating the circum-merger medium \citep{MargalitPiran2020}\,
are unlikely to play a major role. 

\begin{figure*}[!t]
\begin{center}
\hspace*{-0.1in}
\includegraphics[scale=0.5]{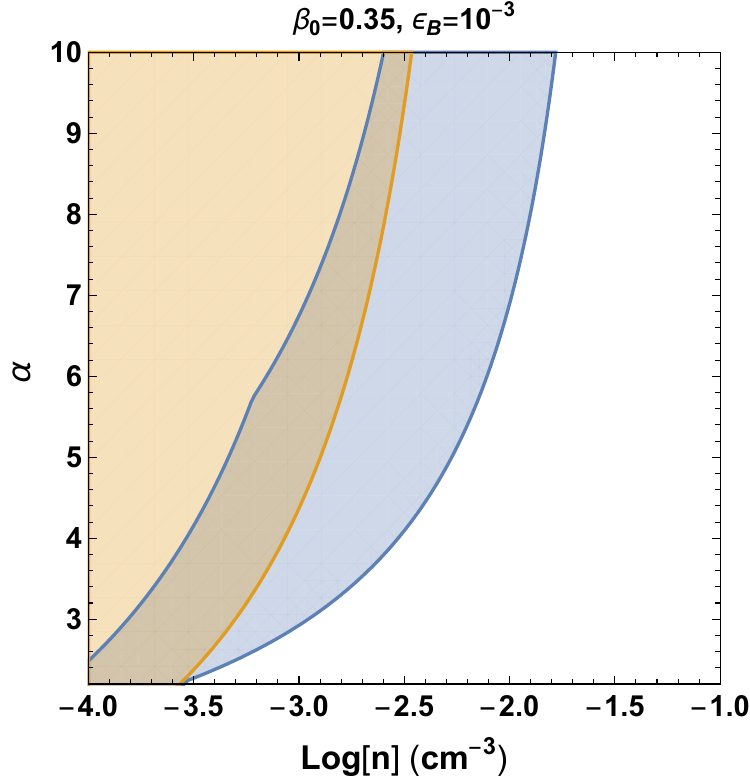}
\includegraphics[scale=0.5]{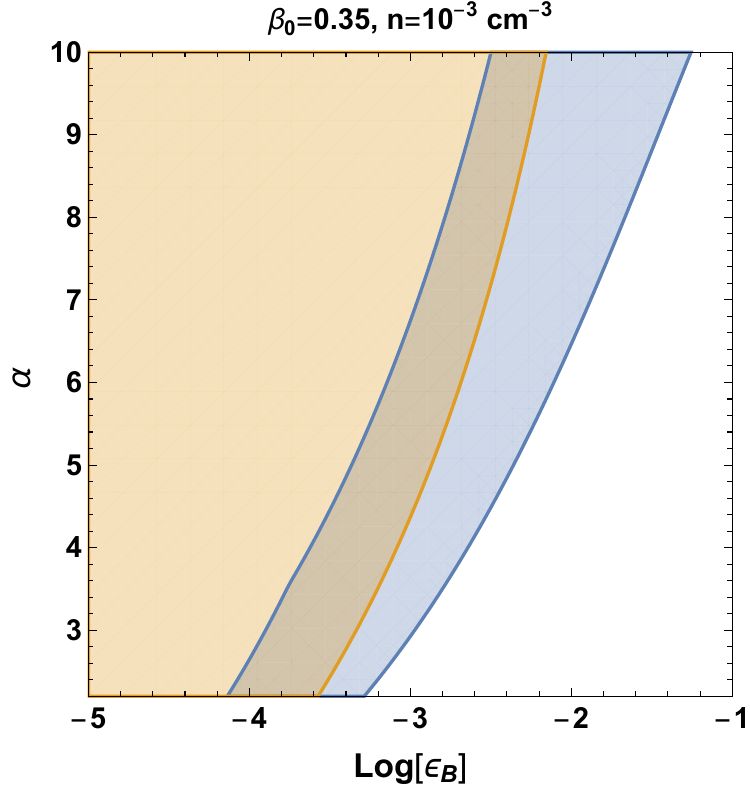}
\includegraphics[scale=0.5]{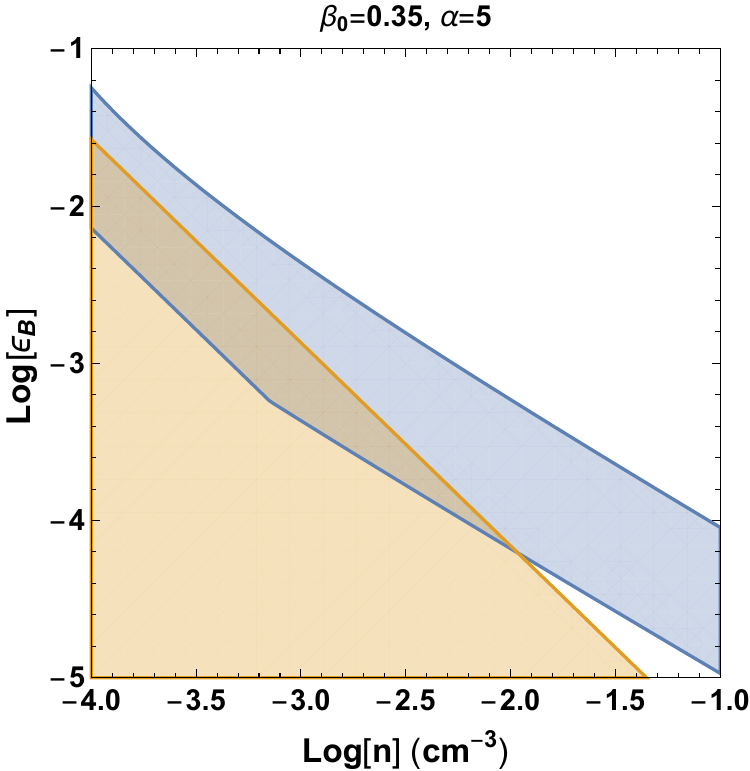}
\includegraphics[scale=0.5]{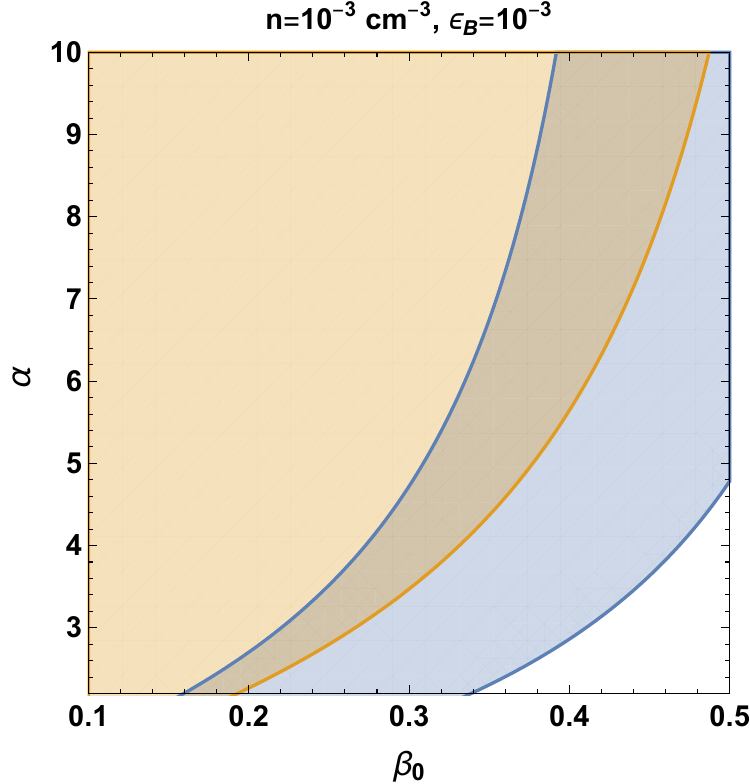}
\vspace*{5mm}\caption{Blue shaded area: region of the parameter space consistent with the X-ray flux excess at 1234 days following the modeling described in \S\ref{Sec:KNmodels}. Orange shaded area: region of the parameter space that is consistent with our radio upper limit at 3\,GHz: $F_{\rm \nu}<5.1\,\rm{\mu Jy}$. The kinetic energy distribution of the kilonova ejecta in the velocity space has been parameterized as $E_{\rm{KN}}\propto (\Gamma \beta)^{-\alpha}$ above $\beta_0$ with $E_{\rm{KN}}(\Gamma_0 \beta_0)=10^{51}\,\rm{erg}$. The shock microphysical parameters adopted in this calculation are $p=2.05$ (consistent with the observational findings of \S\ref{Sec:SpecEvol}) and $\epsilon_{\rm{e}}=0.1$. Two parameters are varied in each plot while the rest are kept fixed to values indicated in the plot title. 
}
\label{Fig:KNAdithan}
\end{center}
\vspace{-0.25cm}
\end{figure*}

\begin{figure*}[!t]
\begin{center}
\hspace*{-0.1in}
\includegraphics[scale=0.42]{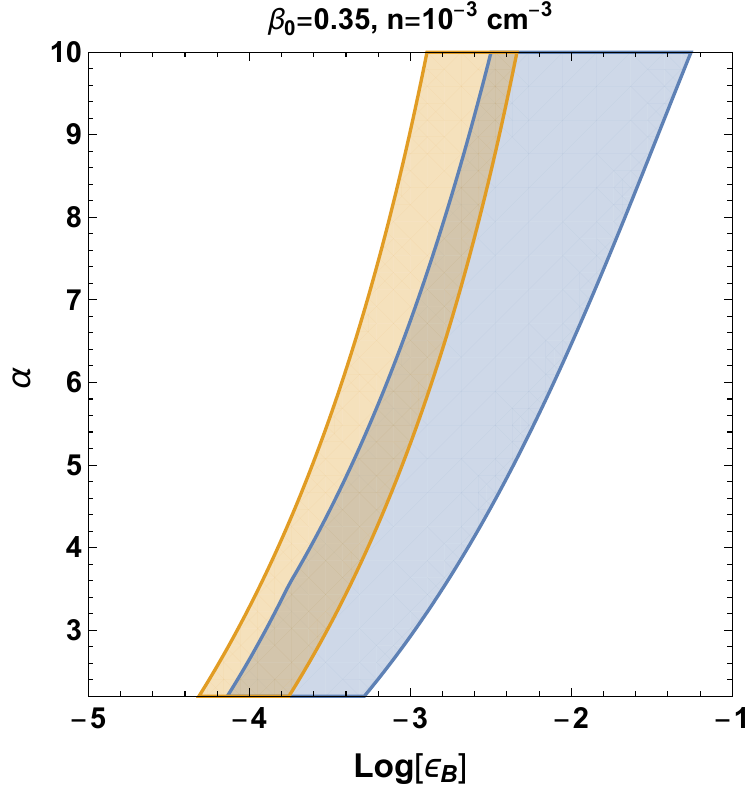}
\includegraphics[scale=0.42]{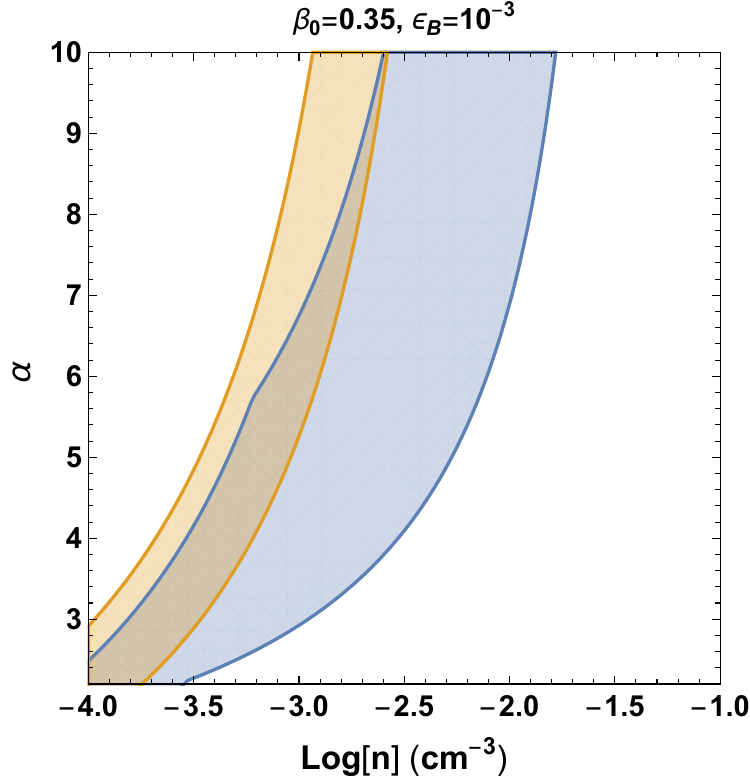}
\includegraphics[scale=0.42]{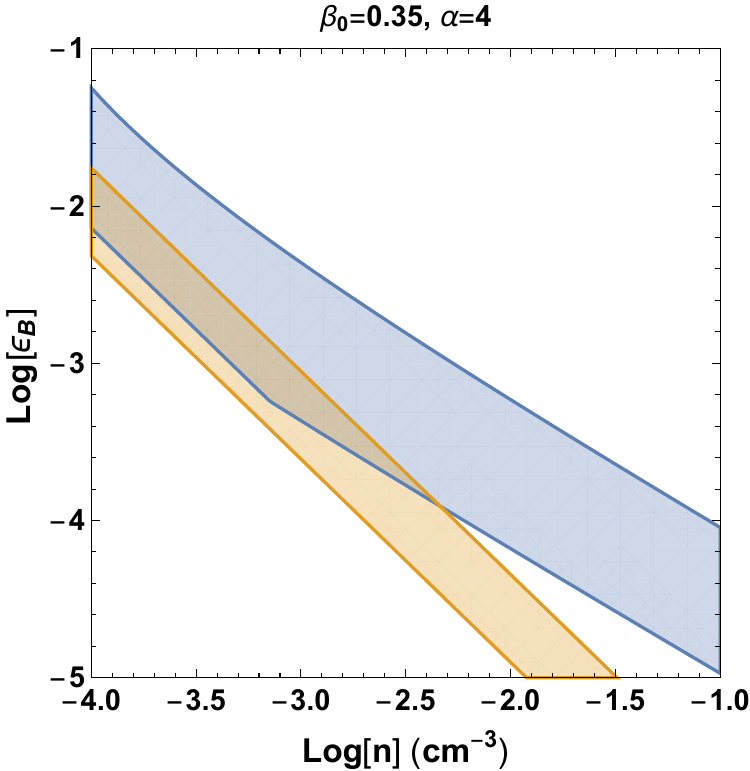}
\caption{Kilonova afterglow parameter space with the same color scheme as Figure \ref{Fig:KNAdithan} where we used the peak pixel flux within one synthesized beam at 3 GHz from \citealt{Balasubramanian21}\, ($F_{\rm \nu}=2.8 \pm 1.3\,\rm{\mu Jy}$) as a constraint on the radio emission from the kilonova. As in Figure \ref{Fig:KNAdithan}, we assume $E_{\rm{KN}}=10^{51}$\,erg, $\epsilon_{\rm{e}}=0.1$, and $p=2.05$.} Our conclusions remain unchanged.
\label{Fig:KNAdithanCaltech}
\end{center}
\end{figure*}

\subsection{Kilonova afterglow models from Kathirgamaraju et al. 2019}\label{SubSec:AdithanKN}
We parameterized the kinetic energy distribution of the kilonova ejecta as a power-law in specific momentum $\Gamma \beta$ for  $\beta > \beta_0$: $E_{\rm{KN}}\propto (\Gamma \beta)^{-\alpha}$ \citep{Kathirgamaraju+2019}. This parameterization captures the properties of the high-velocity tail of all types of kilonova outflows,  including dynamical ejecta and disk winds that might dominate the mass of the blue kilonova component. 
Motivated by the results from the modeling of the thermal emission from the kilonova in the following we adopt  $\beta_0=0.35$ as baseline and a total kinetic energy of $E_{\rm{KN}}(\Gamma_0 \beta_0)=10^{51}\,\rm{erg}$.  We generated a set of multi-wavelength kilonova afterglow light-curves for shock microphysical parameters $p=2.05$ (consistent with the observational findings of \S\ref{Sec:SpecEvol}), $\epsilon_{\rm e}=0.1$, $\epsilon_{\rm B}=[10^{-4}-10^{-2}]$ and circumbinary medium density $n=[10^{-4}-10^{-2}]\,\rm{cm^{-3}}$. As a comparison, studies of the jet afterglow pointed at densities $n>10^{-4}\,\rm{cm^{-3}}$ \citep{MarguttiChornock2021} 
, while multiple studies \citep{Hallinan+2017radiocounterpart,Hajela+2019,Makhathini+2020}\, of the large-scale environment of GW\,170817 at X-ray and radio wavelengths argue in favor of $n\le 10^{-2}\,\rm{cm^{-3}}$. Motivated by the results from  numerical relativity simulations of BNS mergers described below we explore the parameter space for $\alpha=[3-9]$. 

Our results are shown in  Figure \ref{Fig:KNAdithan}, where shaded areas highlight the regions of the parameter space that are consistent with the bright X-ray excess (blue) and the deep radio upper limit (orange). We further show a successful kilonova afterglow model for $\alpha=5$, $n=0.001\,\rm{cm^{-3}}$ and $\epsilon_{\rm B}=0.001$ in Figure \ref{Fig:Xrays}. Consistent with the results from the jet afterglow modeling, current data point to lower density environments with $n<0.01\,\rm{cm^{-3}}$, but otherwise leave the multi-dimensional parameter space largely unconstrained. Specifically, we find that all values of $\alpha=[3,10]$ are consistent with the X-ray and radio data set. This conclusion remains unchanged even if we adopt the peak pixel flux within one synthesized beam at 3 GHz from \cite{Balasubramanian21}\, ($F_{\rm \nu}=2.8 \pm 1.3\,\rm{\mu Jy}$) as a constraint on the radio emission from the kilonova (Figure \ref{Fig:KNAdithanCaltech}).

Using their reduction of the multi-wavelength data set up to $\approx 1200$ days and similar to our preliminary assessment of the properties of the kilonova ejecta properties in \cite{Hajela+2019}, \cite{Balasubramanian21} favor $\alpha \geq 5$ kilonova ejecta profiles assuming a density and the kilonova shock microphysical parameters set to the values of the jet afterglow shock (i.e. n $\sim 10^{-2}\,\rm{cm^{-3}}$, $\epsilon_{\rm{e}} \sim 10^{-2}$, and $\epsilon_{\rm{B}} \sim 10^{-3}$). While for this choice of extrinsic parameters our findings qualitatively agree with the conclusions by \cite{Balasubramanian21}, we note that there is no physical reason for the kilonova shock microphysical parameters to be the same as those of the jet afterglow shock, and relaxing these parameters  leaves the problem unconstrained. Our results are consistent with those from previous analyses that did not include the latest epoch \citep{Hajela+2019,Troja+2020}, and constitute an important advancement with respect to these previous works that were completed before the emergence of a statistically significant new component of emission.


\begin{figure*}[!t]
\begin{center}
\hspace*{-0.1in}
\includegraphics[scale=0.6]{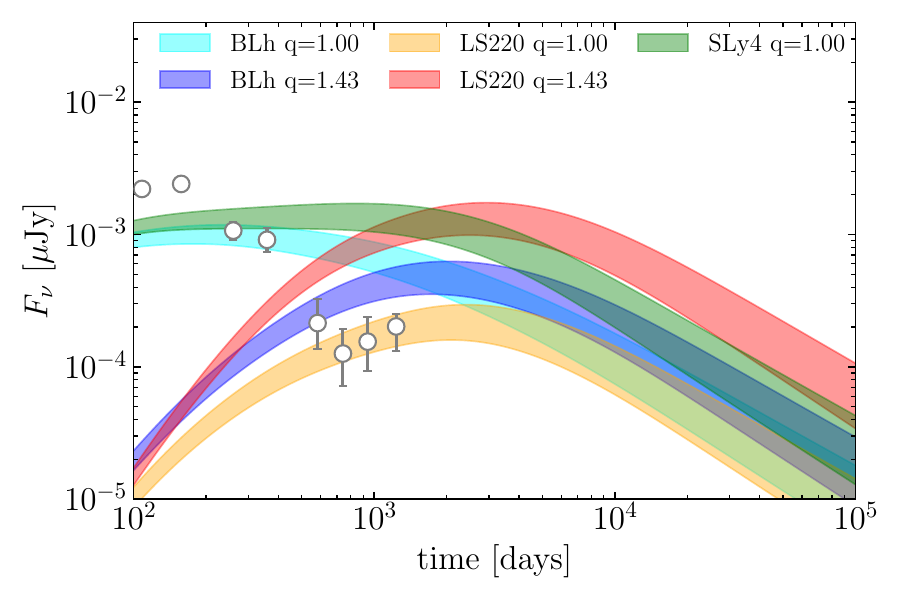}
\includegraphics[scale=0.6]{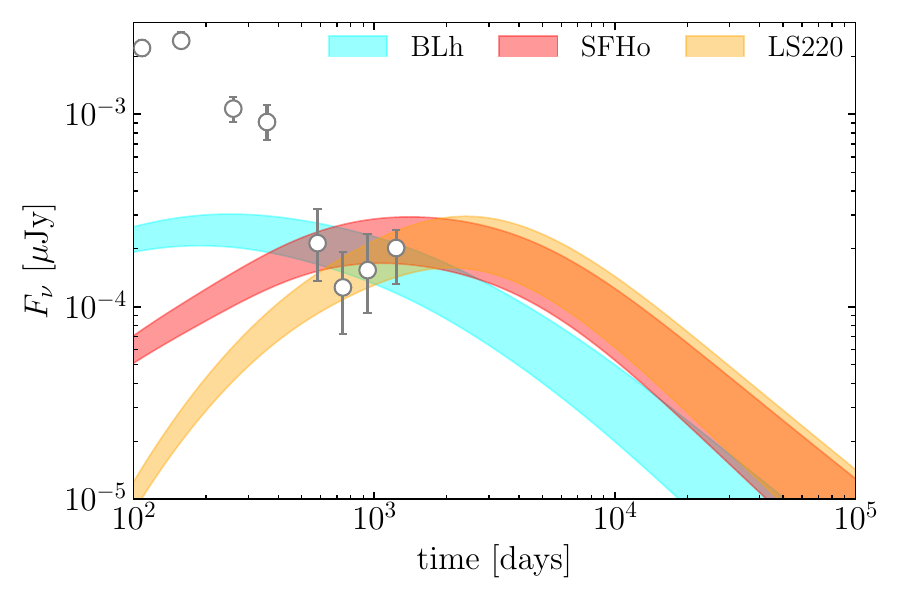}
\caption{\emph{Upper Panel:} Kilonova afterglows from a set of ab-initio numerical relativity BNS merger simulations. In these simulations the kilonova ejecta is of dynamical nature,  with  resulting kinetic energy profiles shown in Figure \ref{Fig:EkGammaBeta}. Different colors correspond to different EoSs (BLh, LS220, and SLy4) and NS mass ratios $q$.  Good quantitative agreement between the numerical relativity predictions and the observation is obtained. 
The light curves are computed assuming an ISM density of $n_{\text{ISM}}=6\times10^{-3}$~cm$^{-3}$, and microphysical parameters, $\epsilon_{\rm e}=10^{-1}$, $\epsilon_{\rm B}=10^{-2}$.
\emph{Lower Panel:} Effect of the extrinsic parameters (i.e. density and shock microphysics) on the kilonova afterglow emission from equal-mass NS binaries (i.e., $q\approx1$ that is typical of the Galactic population) and different EoSs. For LS220, BLh and SFHo current observations are consistent with $n\sim 6\times 10^{-3},5\times 10^{-3},5\times 10^{-3}\,\rm{cm^{-3}}$ and $\epsilon_{\rm B}\sim 10^{-2},2\times 10^{-3},10^{-3}$, respectively, for a fiducial $\epsilon_{\rm e}=0.1$. 
In both panels the viewing angle is assumed to be $30^\circ$ from the polar axis. The bands correspond to light curves with the electron distribution power-law index $p$ varying between $2.05$ and $2.15$.}
\label{Fig:kn_afterglow}
\end{center}
\end{figure*}

\begin{figure*}[!t]
\begin{center}
\includegraphics[scale=0.76]{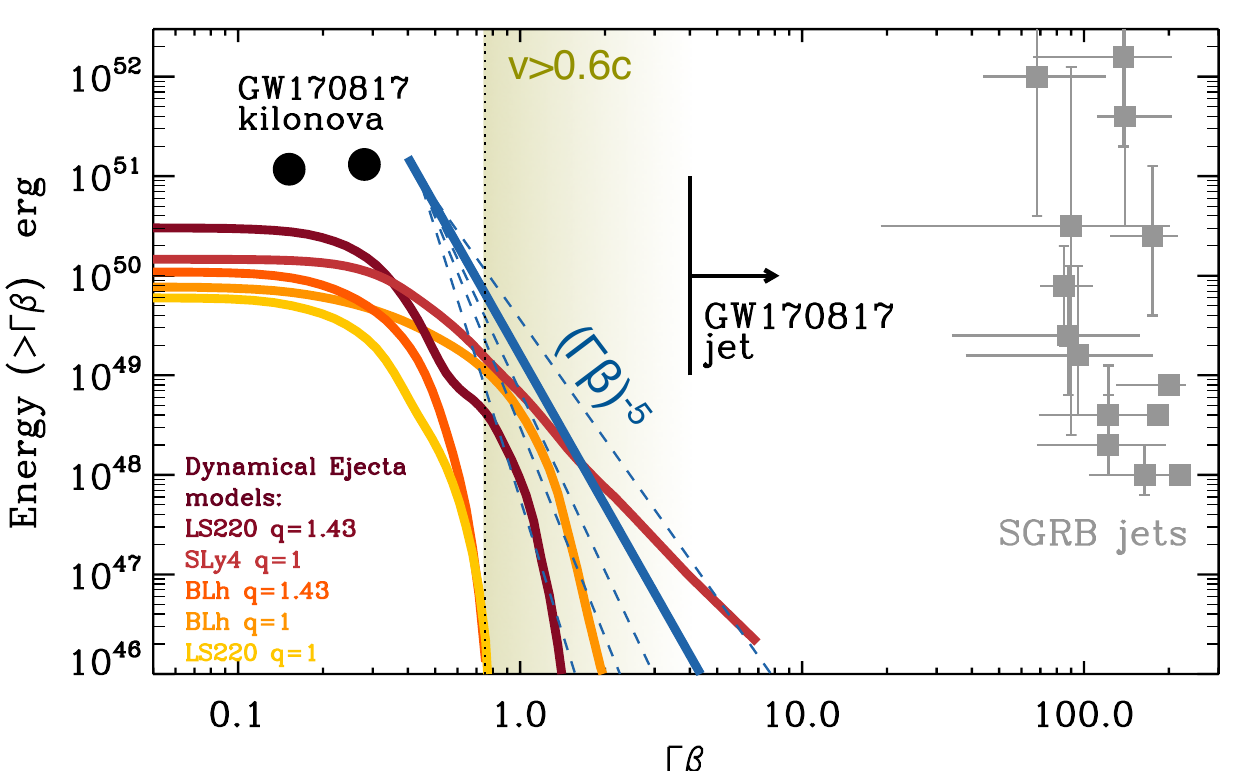}
\includegraphics[scale=0.76]{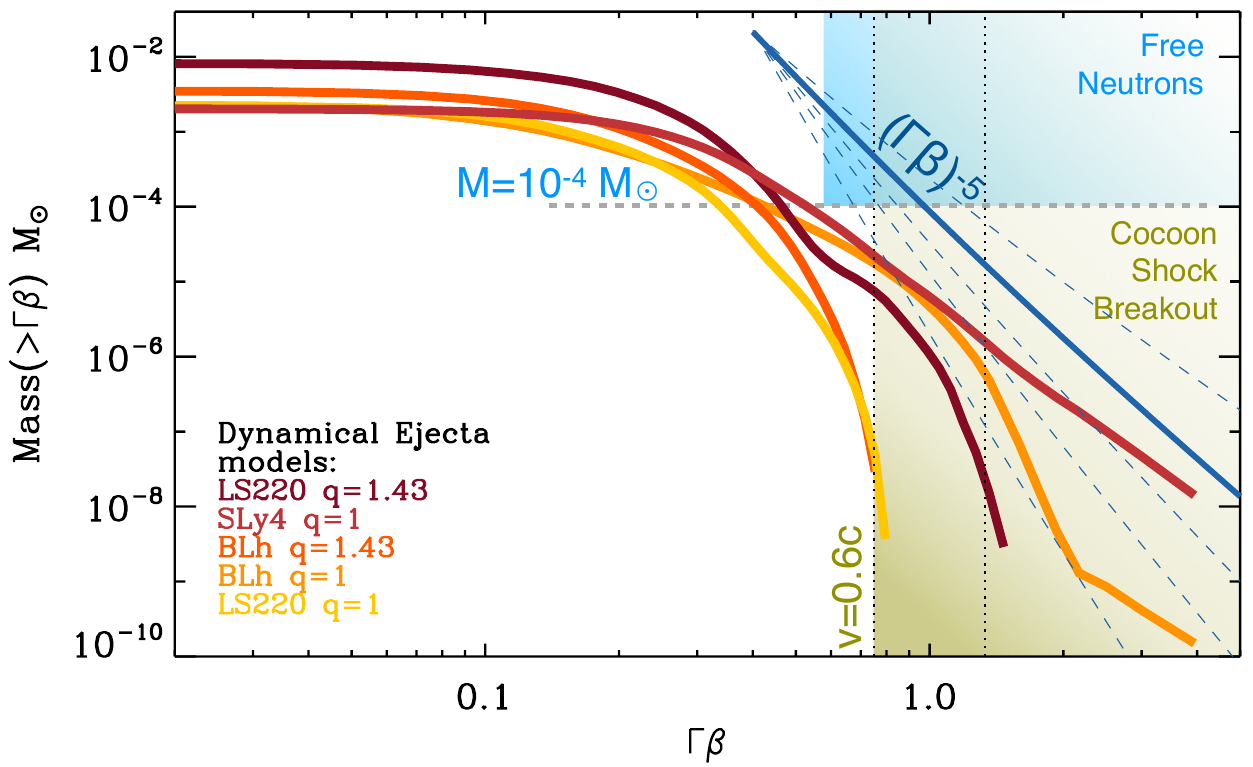}
\caption{
\emph{Upper Panel:} Colored lines: kinetic energy profile of the  fastest kilonova ejecta as a function of specific momentum $\Gamma \beta$. Dark-red to orange shade: dynamical ejecta profiles as inferred from ab-initio numerical-relativity simulations described in \S\ref{Sec:KNmodels} for different EoS and NS mass ratios $q$. Blue lines: $E_{\rm{KN}}(>\Gamma \beta)\propto (\Gamma \beta)^{-\alpha}$ analytical profiles that include the contributions from all types of kilonova ejecta for $\alpha=4,5,6,7,9$. Black filled circles: kinetic energy inferred from the modeling of the UV/optical/NIR kilonova emission \citep{Villar+2017combined}. Grey squares: SGRB jets \citep{WuMacFadyen+2019}. \emph{Lower Panel:} kilonova ejecta profiles in the mass phase-space.  Green colored area: region of the parameter space consistent with a cocoon shock breakout origin of GRB\,170817A \citep{Gottlieb+2017cocoon}. Blue colored area: region of the parameter space which is suggestive of a free-neutron component of the ejecta expected to power a short-lived UV/optical transient.}
\label{Fig:EkGammaBeta}
\end{center}
\end{figure*}

\subsection{Kilonova afterglows from physically-motivated kilonova kinetic energy profiles}\label{SubSec:NRKN}
We consider a set of 76 numerical relativity BNS merger simulations tailored to GW\,170817 \citep{Perego:2019adq, Endrizzi:2019trv, Nedora:2019jhl, Bernuzzi:2020txg, Perego:2020evn, Vsevolod:2020pak}. The simulations were performed using the \texttt{WhiskyTHC} code \citep{Radice+2012,Radice+2014a,Radice+2014b}. The set includes simulations performed at different resolutions and employs five finite-temperature microphysical equations of state (EoSs) that span the (large) range of EoS compatible with current laboratory and astronomical constraints. The simulations self-consistently included compositional and thermal effects due to neutrino emission and re-absorption \citep{Radice:2016dwd, Radice:2018pdn}. The general-relativistic large-eddy simulation (GRLES) method was used to capture subgrid-scale turbulent dissipation and angular momentum transport \citep{Radice:2017zta, Radice:2020ids}.

Dynamical ejecta from these simulations show the presence of a fast moving tail of ejecta, which is 
produced following the centrifugal bounce of the remnant taking place in the first milliseconds of the merger, unless prompt BH formation occurs, in which case there is no bounce \citep{Radice:2018pdn}. The bounce produces a shock wave that is rapidly accelerated by the steep density gradient in the outer layers of the remnant, propels material to trans-relativistic velocities, and propagates into the circumbinary medium. 
Fast moving material could also be accelerated by the thermalization of mass exchange flows between the stars prior to merger \citep{Radice:2018ghv}. However, this alternative scenario typically predicts a faster rise of the synchrotron emission than indicated by observations of GW\,170817. 

The deceleration of this kilonova shock into the medium produces synchrotron radiation. We compute the kilonova synchrotron light curves using the semi-analytic code \texttt{PyBlastAfterglow} \citep{Nedora+2021}.  We have validated this code in the subrelativistic regime by comparing the results it produces using the ejecta profiles from \cite{Radice:2018pdn}, which had been previously analyzed using the code of \cite{Hotokezaka:2015eja} and in the ultra-relativistic regime by comparing our results with those produced by \texttt{afterglowpy} \citep{Ryan+2020}.

 Figure~\ref{Fig:kn_afterglow} collects a representative set of  X-ray light curves for three EoSs (BLh, \citealt{Logoteta:2020yxf, Bernuzzi:2020txg}; LS220, \citealt{Lattimer:1991nc}; and SLy4, \citealt{Douchin:2001sv, daSilvaSchneider:2017jpg}) and two values of the binary mass ratio $q$. This figure highlights the sensitivity of the kilonova afterglow on intrinsic (EoS, $q$) and extrinsic ($n$, $p$, $\epsilon_{\rm e}$, $\epsilon_{\rm B}$) parameters of the binary. 
It is important to emphasize that the overall flux level predicted by our models is strongly dependent on assumed microphysical parameters of the shock. 
However, the light curve temporal evolution only depends on the structure of the ejecta and on the ISM density. Specifically, the peak time of the kilonova emission is of dynamical nature, tracing the deceleration time of the blast wave into the environment \citep{Nakar+2011Nat} and it is thus independent from the parameters that set the level of the emitted flux (like the shock microphysical parameters).

With respect to the intrinsic binary parameters probed by our simulation, we find that binaries which do \emph{not} undergo prompt BH formation are broadly consistent with the observations. Numerical simulations of BNS mergers by \cite{Prakash+2021,Nedora+2021} show  that if prompt collapse to BH occurs in equal mass NS binaries, the kilonova afterglow is expected to be several orders of magnitude fainter than the observed X-ray luminosity of GW\,170817 at $\approx 1000$\,days (e.g. Figure 15 in \citealt{Prakash+2021}). In the case of highly asymmetric NS binaries, the prompt collapse to BH is associated with afterglow light curves that peak at $\approx 10^4$\,days post-merger, which is significantly later than the current epoch (see Figure 4 and 5 in \citealt{Nedora+2021}).
An important conclusion is that prompt BH formation is disfavored \citep{Bauswein:2017,Margalit+2017,Radice:2018jointconstraint},  because the presence of the post-merger bounce appears to be necessary in order to produce sufficient fast and massive outflows to power the kilonova emission. 
Improved higher-resolution targeted simulations are needed to draw more quantitative conclusions.

In addition to the nature of the compact-object remnant, the early detection of a kilonova afterglow a few years after the merger and its future modeling 
can enable fundamental insight into two other still-open questions pertaining to GW\,170817: the presence of a free-neutron component of ejecta, and the origin of the detected prompt $\gamma$-rays \citep{Goldstein17,Savchenko17}. Fast ejecta with mass $\gtrsim 10^{-4}\,\rm{M_{\odot}}$ at velocity $v\ge 0.5$c (light-blue shaded area in Figure \ref{Fig:EkGammaBeta}, lower panel) are expected to lead to a freeze out of the $r$-process \citep{MetzgerFreeNeutrons},  as most neutrons will avoid capture, leaving behind free neutrons that can power a short-lived (i.e. $\approx\,$hrs) but luminous UV/optical transient. Additionally, kilonova ejecta profiles extending to velocities 
$v\ge 0.6\,c$ (light-green shaded area in Figure \ref{Fig:EkGammaBeta}, lower panel)  provide the necessary conditions to produce $\gamma$-rays from a shock breakout of a wide-angle outflow (i.e. the cocoon) inflated by the jet  from the merger ejecta \citep{Bromberg18,Gottlieb+2017cocoon}.  Being sensitive to the presence and properties of the fast kilonova ejecta, the kilonova afterglow is thus  a  probe of the merger dynamics and nature of the compact object remnant.

We conclude by remarking that a general, robust and testable prediction of the kilonova afterglow models is that of a persistent source of emission across the electromagnetic spectrum, which is not expected to become fainter for thousands of days, and might even become brighter during this period of time. Eventually, the kilonova afterglow will appear as a detectable source in the radio sky and might even be detectable via deep optical observations from space.

\section{Emission from a Compact-Object Remnant}
\label{Sec:Engine}
An alternative explanation of rising X-rays without accompanying bright radio emission is that of central-engine powered radiation, i.e. radiation powered by an energy release associated with the compact-object remnant either in the form of accretion (for a BH remnant) or spin-down energy (for a long-lived NS remnant). The nature of the compact-object remnant of GW\,170817 is a fundamentally open question that directly relates to the NS EoS. While post-merger GWs were inconclusive, the observational evidence for (i) a blue kilonova component associated with a large mass of lanthanide-free ejecta and kinetic energy $\approx 10^{51}\,\rm{erg}$ \citep{Evans+2017,Villar+2017combined,Cowperthwaite+2017EM,Nicholl+2021,Bulla+2019}, and (ii) the uncontroversial evidence for a successful relativistic jet \citep{Alexander+2018,Mooley+2018superluminal,Ghirlanda+2019} together with energetics arguments strongly disfavor either a prompt collapse to a BH or a long-lived NS remnant. These arguments and observations argue in favor of a hypermassive NS that collapsed to a BH within a second or so after the merger \citep{Granot+2017,Margalit+2017,Shibata+2017,Metzger+2018,Rezzolla+2018,Gill+2019,Ciolfi2020,Murguia-Berthier+2020}.  
While the most likely scenario is that of a BH remnant at the current time of the observations, in the following we also consider the less-likely case of a spinning-down NS for completeness (see however \citealt{Piro+2019}).

\subsection{Accreting BH remnant scenario}
\label{SubSec:BHaccretion}
The Eddington luminosity for accretion onto a remnant BH of mass $M_{\bullet} \sim 2.5M_{\odot}$\,\citep{ligopropgw170817}  
 of GW\,170817 is given by
\begin{equation}
\label{Eq:Ledd}
L_{\rm Edd} = \frac{4\pi G M_{\bullet}c}{\kappa_{\rm es}} \approx 8\times 10^{38}\left(\frac{M_{\bullet}}{2.5M_{\odot}}\right)\,{\rm erg\,s^{-1}}, 
\end{equation}
where $\kappa_{\rm es} = Y_e\sigma_{\rm T}/m_p \approx 0.16$ cm$^{2}$ g$^{-1}$ is the approximate electron scattering opacity for fully ionized matter comprised of heavy elements (electron fraction $Y_{e} \simeq 0.4$).

From hydrodynamical simulations of BNS mergers, the rate of fall-back accretion is $\dot{M}|_{t_0} \sim 2\times 10^{-4}M_{\odot}$ s$^{-1}$ on a timescale of $t_0 \sim 1$ s after the merger \citep{Rosswog2007}.  A more important source of fall-back material may arise from the accretion disk outflows \citep{Fernandez+2013}, which likely dominated the kilonova ejecta in GW\,170817 \citep{Radice2020review}.  
If a few tens of percent of the total ejecta mass $\approx 0.06M_{\odot}$ inferred for  GW\,170817 \citep{Villar+2017combined,Cowperthwaite+2017EM,Nicholl+2021,Arcavi_2018,Waxman+2018}  
were to fall back to the BH on a timescale comparable to the predicted accretion disk lifetime $\sim 1$ s, the mass fall-back rate would be orders of magnitude higher, $\dot{M}|_{\rm t_0} \sim 10^{-2}M_{\odot}$ s$^{-1}$.

Based on the expectation that $\dot{M} \simeq \dot{M}_{t_{\rm fb}}(t/t_{\rm 0})^{-5/3}$ at times $t \gg t_{0}$ for marginally bound material \citep{Rees1988}, the total X-ray accretion luminosity is given by
\begin{eqnarray}
L_{\rm X} &\approx& \frac{\eta}{f_{\rm b}} \dot{M}c^{2}  \approx 10^{39} \rm{\,erg\,s^{-1}}\left(\frac{f_{\rm b}}{0.1}\right)^{-1} \nonumber \\
&\times&\left(\frac{\eta}{0.1}\right)\left(\frac{\dot{M}\vert_{t_0}}{10^{-2}M_{\odot}\,\rm s^{-1}}\right)\left(\frac{t}{1000{\,\rm days}}\right)^{-5/3},
\label{eq:Laccretion}
\end{eqnarray}
where the radiative efficiency $\eta$ has been normalized to that of a thin disk orbiting a BH of dimensionless spin $a \approx 0.6-0.8$ \citep{Novikov1973}, as expected for the remnant of a BNS merger.  Here $f_{\rm b}$ is the geometric beaming fraction of the X-ray emission. We expect  $f_{\rm b}\ll 1$ for sources at or near the Eddington luminosity (e.g., Ultraluminous X-ray sources, ULXs, \citealt{Walton+2018} 
) due to powerful disk outflows that generate a narrow accretion funnel \citep{King2009}.  We have normalized $f_{\rm b}$ to a lower limit based on the observer's viewing angle \cite{Mooley+2018superluminal,Hotokezaka+2019,Ghirlanda+2019}\,
$\theta_{\rm obs} \approx 0.4$ with respective to the original binary axis ($\simeq$ accretion disk angular momentum axis): $f_{\rm b,min} \approx \theta_{\rm obs}^{2}/2 \sim 0.1$.

\begin{figure}[!t]
\begin{center}
\scalebox{1.}
{\includegraphics[scale=0.5]{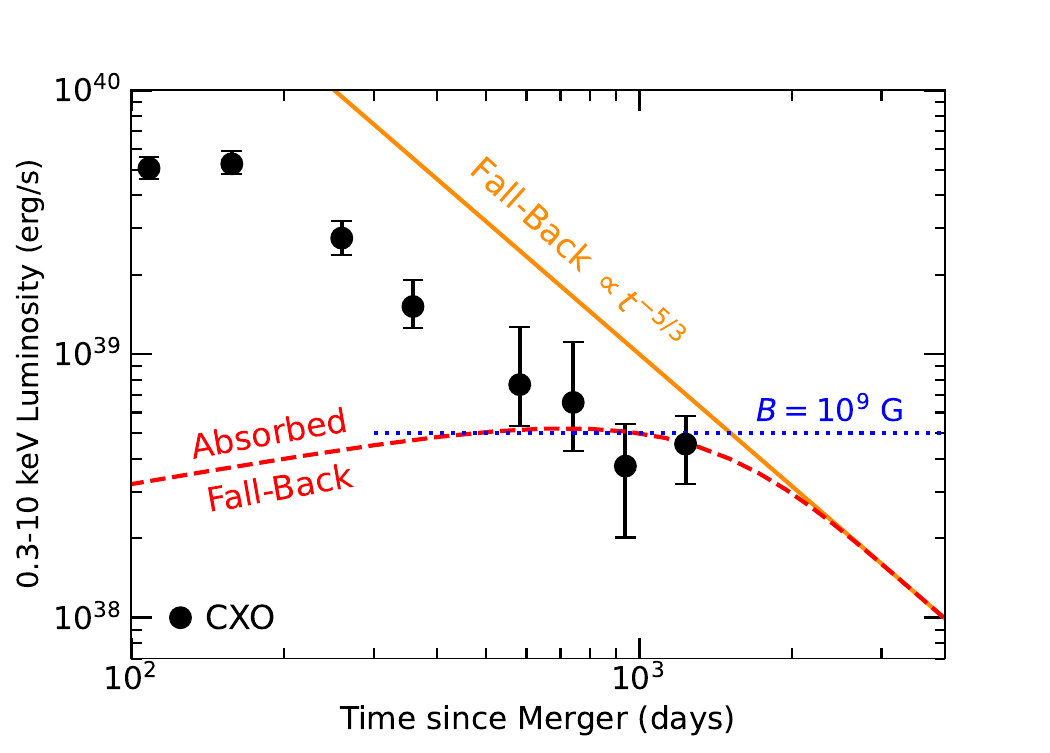}}
\caption{Observed $0.3$ -- $10$\,keV X-ray luminosity (black filled circles) compared to two sources of energy to power the X-ray excess in the compact-object powered scenario:  (1) accretion-powered fall-back luminosity, both intrinsic (orange solid line) and observed (red dashed line), i.e. with a correction for absorption by the kilonova ejecta of the form $\propto (1-e^{-(t/t_{\rm thin})^{2}})$, where $t_{\rm thin} \approx 1000$ days (Eq.\ref{eq:thin}). And, (2) magnetar spin-down luminosity (Eq. \ref{eq:Lsd2}, dotted blue line) for $B \sim 10^{9}$\,G to match the level of the observed X-ray emission.}
\label{Fig:magnetar}
\end{center}
\vspace{-0.75cm}
\end{figure}

In analogy with X-ray binaries in the ``ultra luminous" state \citep{Gladstone+2009} the spectra of stellar mass BHs accreting close to the Eddington rate are satisfactorily modeled by a thermal accretion disk plus power-law component with a high-energy exponential break.
Ignoring relativistic terms and color corrections, the effective temperature of the disk emission can be estimated as
\begin{eqnarray}
2\pi R_{\rm isco}^{2}\sigma T_{\rm eff}^{4} = f_{\rm b}L_{\rm X},
\end{eqnarray}
where $R_{\rm isco} \approx 3GM_{\bullet}/c^{2}$ is the innermost radius of the disk for a BH of spin $a \approx 0.6-0.8$.  This gives
\begin{eqnarray}
kT_{\rm eff} &\simeq& 2\,{\rm keV} \left(\frac{f_{\rm b}}{0.1}\right)^{1/4} \nonumber \\
&\times& \left(\frac{L_{\rm X}}{5\times 10^{38}{\rm erg\,s^{-1}}}\right)^{1/4}\left(\frac{M_{\bullet}}{2.5M_{\odot}}\right)^{-1/2},
\label{eq:Teff}
\end{eqnarray}
i.e. in the range of the {\it CXO} sensitivity window for the observed $L_{\rm X} \approx 5\times 10^{38}{\rm erg\,s^{-1}}$ at 1234 days ( Table \ref{Tab:xrayanalysistab}).

We now consider the question of the observability of this X-ray emission. The X-ray rise time will be determined by the maximum of two timescales.  The first is the timescale for the accretion rate to drop sufficiently that the beaming fraction $f_{\rm b} \propto (\dot{M}/\dot{M}_{\rm Edd})^{-2} \propto t^{10/3}$ \citep{King2009} increases to the point that  the angle of the accretion funnel $\theta_{\rm b} \propto f_{\rm b}^{1/2} \propto t^{5/3}$ enters the observer's viewing angle $\theta_{\rm obs} \approx 0.4$.  Given that $L_{\rm X}$ at the present epoch is $\lesssim L_{\rm Edd}$ (Eq. \ref{Eq:Ledd}), we conclude that this effect may still play a role in generating a rising X-ray luminosity.

A second timescale for the X-rays to be able to reach the observer is that required for the kilonova ejecta to become transparent to the X-rays.  Assuming that the $r$-process ejecta have a bound-free opacity to photons of energy $\sim 1$ keV which is similar to that of iron group elements $\kappa_{\rm X} \approx 10^{4}$ cm$^{-2}$ g$^{-1}$, this will take place after a time

\begin{eqnarray}
t_{\rm thin} &=& \left(\frac{3M_{\rm ej}\kappa_{\rm X}}{4\pi v_{\rm ej}^{2}}\right)^{1/2} \nonumber \\ 
&\approx& 2000\,{\rm days}\,\left(\frac{v_{\rm ej}}{0.1{\rm c}}\right)^{-1} \nonumber\\
&\times& \left(\frac{\kappa_{\rm X}}{10^{4}\,\rm cm^{2}\,g^{-1}}\right)^{1/2}\left(\frac{M_{\rm ej}}{0.06M_{\odot}}\right)^{1/2},
\label{eq:thin}
\end{eqnarray}

where we have normalized the ejecta mass $M_{\rm ej}$ and velocity $v_{\rm ej}$ to characteristic values for the (dominant) red/purple ejecta component inferred by modeling the optical/IR kilonova of GW\,170817 \citep{Villar+2017combined,Cowperthwaite+2017EM,Drout+2017LC,Kilpatrick+2017}. 

Given that the ejecta density may be lower than average for our high altitude viewing angle $\theta_{\rm obs} \approx 0.4$, and hence $t_{\rm thin}$ somewhat over-estimated, we conclude that $t_{\rm thin}$ is also likely to be comparable to the present epoch. Figure \ref{Fig:magnetar} shows the evolution of the accretion-powered fall-back X-ray luminosity on a BH remnant, both intrinsic (orange solid line) and observed (red dashed line), i.e. with a correction for absorption by the kilonova ejecta of the form $\propto (1-e^{-(t/t_{\rm thin})^{2}})$, where we used $t_{\rm thin} \approx 1000$ days, as the time when the ejecta becomes optically thin. An absorption cause for the X-ray rise could in principle be tested by a strong suppression of soft X-ray photons due to the rapidly increasing bound-free opacity towards lower-energy X-rays.  However, due to faintness of the X-ray source (which leads to very low-count statistics, \S\ref{SubSec:CXOspecanalysis}) combined with the progressive loss of sensitivity of the \emph{CXO} at soft X-ray energies, this effect cannot be tested at present with any statistically meaningful confidence.

One potential constraint on this scenario comes from earlier IR/optical observations, since at earlier epochs the absorbed X-rays would be reprocessed to IR/optical radiation.  For instance, to explain $L_{\rm x} \sim 5\times 10^{38}$ erg s$^{-1}$ at $t_{\rm now} \sim 10^{3}$ days, the accretion power on a timescale of $t_{\rm KN} \sim 1$ week after the merger would be higher by a factor $\sim (t_{\rm now}/t_{\rm KN})^{5/3}\approx 4000$, or $\sim 2\times 10^{42}$ erg s$^{-1}$. The bolometric UV/optical/IR emission \citep{Cowperthwaite+2017EM,Arcavi_2018,Waxman+2018}  from the kilonova of GW\,170817 reached $L\approx 10^{41}\,\rm{erg\,s^{-1}}$. The accretion power would thus exceed the bolometric output of the kilonova on this timescale by a factor $\gtrsim 10$.  Even more stringently, extrapolating back to the last \emph{HST} optical detection of GW\,170817 at $\approx 360$ days since merger leads to values $\approx 10^2$ times larger than the observed \emph{HST} luminosity. At 360 days the optical flux density inferred from \emph{HST} observations is perfectly consistent with the power-law spectrum that extends from the radio band to the X-rays \citep{Fong+2019} and it is thus dominated by jet-afterglow emission. 

However, there are two effects that act to alleviate these constraints.  Firstly, at these earlier epochs the fall-back rate is highly super-Eddington.  The efficiency with which the fall-back material reaches the central black hole may be drastically reduced at these early times due to the inability of the super-Eddington accretion to radiatively cool \citep{Rossi+2009}.  Furthermore, the radiative efficiency $\eta$ of highly super-Eddington accretion flows may be substantially reduced relative to the near or sub-Eddington accretion rate which characterizes the present epoch.  Finally, it is unclear if most of the reprocessed power will emerge in the optical/NIR bands; if lanthanide-series atoms dominate the cooling of the gas in the nebular phase then much of the reprocessed emission may emerge in the mid-IR bands \citep{Hotokezaka+2021}.  On the other hand, {\it Spitzer} observations \citep{Villar+2018, Kasliwal+2019spitzer} 
revealed the 4.5$\mu$m luminosity to be $\sim 10^{38}$ erg s$^{-1}$ on a timescale $\approx 74$ days after the merger, at which time the fall-back accretion rate would be a factor $\sim 100$ higher than at present epoch.  Thus we conclude that the reprocessing into the IR band is not a viable option, and would have to rely instead on the reduced accretion efficiency of the fall-back material onto the BH.

We end by commenting on the expected broadband spectrum.
If the GW\,170817 remnant is accreting at or close to the Eddington limit, it is valuable to contrast its observational properties with those of the ultra-luminous X-ray sources (ULXs), which accrete at or above the Eddington limit for compact objects at $\sim1\,M_{\odot}$.
Radio observations of ULX sources place upper limits on the radio power of $L_{\rm R} \lesssim 10^{24}\,\rm{erg\,s^{-1}\,Hz^{-1}}$ \citep{Kording+2005}, corresponding to a flux density limit of $\lesssim 1\,\rm{\mu Jy}$ at the distance of GW\,170817, which is below the level of our latest radio upper limit of $\approx 5\,\rm{\mu Jy}$ (3$\times$RMS, \S\ref{SubSec:VLA_obs}) and  comparable to the local image RMS in our deep VLA observations at 3\,GHz. The lack of a radio counterpart of GW\,170817 is consistent with observations of XRBs in the ``soft'' state, which can accrete at a significant fraction of the Eddington rate and have no associated persistent radio emission \citep{Fender+2004}. Similarly, if GW\,170817 is accreting in a ``hard'' state (associated with an X-ray spectrum peaking at higher energies compared to the soft state), where the X-ray and radio emission are strongly coupled \citep{Corbel+2003}, we would only expect a radio flux density of $\sim10^{22}\,\rm{erg\,s^{-1}\,Hz^{-1}}$ based on our measured \emph{CXO} luminosity and the radio X-ray correlation derived from an ensemble of 24 X-ray binaries in the hard state \citep{Gallo+2014}. Typically, X-ray binaries are only in the hard state while in quiescence (accreting at some small fraction of the Eddington rate) or while in outburst where they typically make the hard to soft state transition \citep{Dunn+2010} at around $\sim0.01\,L_{\rm{Edd}}$ to $\sim0.1\,L_{\rm{Edd}}$. However, high X-ray luminosity hard states have been observed in the XRB GRS 1915$+$105 \citep{Rushton+2010, Motta+2021}, but the associated radio emission would still be well below our detection threshold.
We conclude by emphasizing that a solid expectation from this scenario is that of a different radio-to-X-ray spectrum than the jet afterglow, with less luminous radio emission than expected based on the jet-afterglow spectral slope. This is consistent with our observational findings (\S\ref{Sec:SpecEvol}). Differently from the kilonova afterglow (\S\ref{Sec:KNmodels},  Figure \ref{Fig:kn_afterglow}), in the BH fall-back accretion scenario the X-ray luminosity is expected to to be continuously decreasing
with time ( Figure \ref{Fig:magnetar}). 

To conclude, an accretion-powered origin of the emerging component of X-ray emission is a potentially viable explanation and would naturally account for the broadband spectrum \emph{if} the efficiency of the super-Eddington fall-back matter reaching the black hole is suppressed sufficiently to prevent the accretion luminosity from violating the observed kilonova luminosity at earlier times.\footnote{We note that a similar scenario has been proposed by \cite{Ishizaki+2021fallback}, which was released a few days after a first version of this paper appeared on the arXiv.} This scenario is further supported by $\alpha$-viscosity hydrodynamical simulations presented in \citet{Metzger+2021}.

\vspace{-0.2cm}\subsection{Spinning-down magnetar scenario}\label{SubSec:magnetar}
Alternatively, the additional X-ray component could be powered by spin-down energy from a long-lived magnetar remnant\footnote{We note that the thermal X-ray luminosity of a cooling NS at this epoch is expected to be $\lll L_{\rm{X,obs}} \approx 5 \times 10^{38}\,\rm{erg s^{-1}}$ (see Figure 9 in \citealt{Beznogov+2020}).} (\citealt{Piro+2019, Troja+2020},see however \citealt{Radice+2018longlived}). While there are theoretical arguments against the long-lived magnetar remnant  scenario \citep{Margalit+2017}, we consider this scenario here for completeness.  

The massive NS remnant created by a BNS merger will in general have more than sufficient angular momentum to be rotating near break-up \citep{Radice+2018longlived}. A NS of mass $M_{\rm ns}$ rotating near its mass-shedding limit possesses a rotational energy
\begin{eqnarray}
E_{\rm rot} = \frac{1}{2}I\Omega^{2} &\simeq& 1\times 10^{53}\left(\frac{I}{I_{\rm LS}}\right) \nonumber\\
&\times&\left(\frac{M_{\rm ns}}{2.5 M_{\odot}}\right)^{3/2} \left(\frac{P}{\rm 0.7\rm ms}\right)^{-2}\,{\rm erg},
\label{eq:Erot}
\end{eqnarray}
where $P = 2\pi/\Omega$ is the rotational period and $I$ is the NS moment of inertia, which we have normalized to an approximate value for a relatively wide class of nuclear equations of state $I_{\rm LS} \approx  1.3\times 10^{45}(M_{\rm ns}/1.4M_{\odot})^{3/2}\mathrm{\ g\ cm}^{2}$ \citep{Lattimer+2005}.

The spin-down luminosity $L_{\rm sd}$ of an aligned dipole rotator of surface field strength $B$ with $I = I_{\rm LS}$ is \citep{Philippov+2015}\, 
\begin{eqnarray}
L_{\rm sd}   &=&  7\times 10^{50}\,{\rm erg\,s^{-1}}\left(\frac{B}{10^{15}\,{\rm G}}\right)^{2}\nonumber\\
&\times&\left(\frac{P_{\rm 0}}{\rm 0.7\,ms}\right)^{-4}\left(1 + \frac{t}{t_{\rm sd}}\right)^{-2}
\label{eq:Lsd}
\end{eqnarray}
where we have taken $R_{\rm ns} = 12\,{\rm km}$ as the NS radius, and 
\begin{eqnarray}
t_{\rm sd} = &\left.\frac{E_{\rm rot}}{L_{\rm sd}}\right|_{t = 0} & \nonumber \\
&\simeq 150\,{\rm s}\left(\frac{I}{I_{\rm LS}}\right)&\left(\frac{B}{10^{15}\,{\rm G}}\right)^{-2}\left(\frac{P_{\rm 0}}{\rm 0.7\,ms}\right)^{2}
\label{eq:tsd}
\end{eqnarray}
is the characteristic spin-down time over which an order unity fraction of the rotational energy is removed, where $P_{0}$ is the initial spin-period and we have assumed a remnant mass of $M = 2.3M_{\odot}$.

The natural spin-down timescale, $t_{\rm sd}$, of $\sim150$ seconds (Equation \ref{eq:tsd}), is $\sim6$ orders of magnitude shorter than the observed $\sim1000$ day timescale for the emergence of excess X-ray emission. Accommodating $t_{\rm sd}$ to much-increased $\sim1000$ day timescale implies an a-priori unlikely reduction in the magnetic field, an increase of the initial spin period, or both.
From Eq. \ref{eq:Lsd}:
\begin{eqnarray}
L_{\rm sd} &\simeq& \nonumber\\
&7&\times 10^{50}\,{\rm erg\,s^{-1}}\left(\frac{I}{I_{\rm LS}}\right)\left(\frac{B}{10^{15}\,{\rm G}}\right)^{2}\left(\frac{P_{\rm 0}}{\rm 0.7\,ms}\right)^{-4}
\label{eq:Lsd2}
\end{eqnarray}
Matching the observed excess X-ray luminosity $L_{\rm X} \sim 5\times 10^{38}$ erg s$^{-1}$ would require an extremely weak magnetic field, $B \sim 10^{9}$ G. While this value is in the range of B inferred for recycled pulsars, this magnetic field is much smaller than the field strength $\gtrsim 10^{16}$ G expected to be amplified inside the remnant during the merger processes \citep{Kiuchi+2015}. 
The calculations above do not include the effects related to gravitational-wave losses that have been proposed in the context of the  long-lived NS remnant scenario to dominate the magnetar spin-down at early times to avoid violating the inferred kilonova energy.  
However, it would still require fine-tuning to match $L_{\rm sd}$ to the observed $L_{\rm X}$ for a more physical value of $B$. Furthermore, unlike the BH case (Eq.~\ref{eq:Teff}), there is no reason {\it a priori} to expect the magnetar emission to be largely confined to the X-ray range.

\section{SUMMARY AND CONCLUSIONs}\label{Sec:summary}
We presented the results from our coordinated \emph{CXO}, VLA and MeerKAT campaign of GW\,170817 at $\delta t = 900$ -- $1273$ days (March 2020 to February 2021). Our observations are public and have been partially presented by \cite{Troja+2020} (for data at $\delta t < 950$ days), \cite{Balasubramanian21}, and \cite{Troja+2021}. Our X-ray observations at $\delta t = 940$ and $1234$ days provide the first evidence for a statistically significant deviation from the off-axis jet model and the emergence of a new X-ray component of emission.\footnote{We note that the $\delta t=940$ day dataset would not on its own establish a statistically-significant excess over prior extrapolations.} Our detailed observational findings can be summarized as follows:

\begin{itemize}
\item We found evidence for bright X-ray emission from GW\,170817 with a statistical significance of $7.2\,\sigma$ (\S\ref{SubSec:CXOsrcctrate}) at $\delta t\approx1234$ days with luminosity of $\sim 5 \times 10^{38}\,\rm{erg\,s^{-1}}$. 
This emission is a factor $\approx 4$  larger than the extrapolation of the structured-jet model to the present epoch (Figure \ref{Fig:Xrays}). We employed two independent approaches to estimate the statistical significance of the X-ray excess. For both approaches the statistical tests are performed in the count-rate phase space to minimize the role of any effect related to the flux calibration and self-consistently account for the Poisson nature of the process. The first approach utilizes multi-wavelength jet afterglow light-curves generated with \texttt{JetFit}, while the second approach is jet-model agnostic and adopts an achromatic simple power-law flux decay. Based on these two independent tests  we conclude that the \emph{CXO} observations at $\delta t>900$ days support the evidence of an excess of X-ray emission  compared to the  predictions from the earlier broad-band evolution with statistical significance in the range $3.1\,\sigma$ -- $3.9\,\sigma$ 

\item In contrast to the X-rays, we find no evidence for significant radio emission at the location of GW\,170817 (Figure \ref{Fig:Xrays}, lower panel,  and Figure \ref{Fig:xrayradioimg}), and we place $3\,\sigma$ flux density upper limits of $39$, $5.1$, and $5.1$ $\mu$Jy at mean frequencies of $0.8$, $3$ and $15$\,GHz, respectively, with MeerKAT and the VLA ($3\times \rm{RMS}$,  \S\ref{SubSec:VLA_obs} and \S\ref{SubSec:MeerKAT_obs}).  
\item While there is no evidence for X-ray spectral evolution using the X-ray data alone, the lack of detectable radio emission at the time of the X-ray excess suggests hardening of the non-thermal emission from GW\,170817 (Figure \ref{Fig:SED}) compared to jet afterglow 
with a statistical confidence $\ge 92\%$ -- $99.2\%$ ( Figure \ref{Fig:SpecEvol1234days}). Therefore, these results suggest the evolution of the radio-to-X-ray broadband spectrum towards lower values of $p$ (where the spectrum is $F_{\rm \nu}\propto \nu^{-(p-1)/2}$) and constitute the first indication of spectral evolution of the non-thermal emission from GW\,170817 (Figure \ref{Fig:SED}).
The radio flux density recently reported by \cite{Balasubramanian21} further strengthens these conclusions (Figure \ref{Fig:SpecEvol1234days}).
\end{itemize}

A number of factors could in principle lead to a late-time X-ray light-curve flattening as the observations suggest. 
We discuss the late-time evolution of the jet 
in \S\ref{Sec:latetimejet} as one of the potential scenarios and conclude that to explain the excess of emission it 
would require an ad hoc evolution of key physical parameters of the system 
and is thus disfavored. We propose two alternative explanations: (i) the emergence the kilonova afterglow; (ii) emission from accretion processes on the compact-object remnant.

The emergence of the kilonova afterglow, which originates from a quasi-spherical shock that is different from the jet afterglow shock, can naturally explain the observed broadband spectral evolution of the radiation, as the value of $p$ may be different in the two shocks. In this context the lower value of $p$ suggested by our observations is consistent with the expectations from the 
the theory of Fermi acceleration in the test particle limit \citep{Bell1978,BlandfordOstriker1978,BlandfordEichler1987}\,  for mildly relativistic shocks, such as that produced by the kilonova. 
From our exploration of the kilonova afterglow emission with analytical kinetic energy profiles ($E_{KN}(\Gamma \beta)\propto (\Gamma\beta)^{-\alpha}$, \S\ref{SubSec:AdithanKN}) and physically-motivated $E_{KN}(\Gamma\beta)$ (\S\ref{SubSec:NRKN}, Figure \ref{Fig:EkGammaBeta}) we find that 
ejecta profiles with $\alpha=4$ -- $6$ can reasonably account for observations at $\delta t > 900$ day (Figure \ref{Fig:Xrays}).
However, as discussed in \S\ref{SubSec:AdithanKN}, the parameter space is currently poorly constrained (Figure \ref{Fig:KNAdithan}). 
Similarly, we find that a variety of NS EoS and binary mass ratios can accommodate our observations (also see \citealt{Nedora+2021}). However, a common ingredient of successful models is  binaries that do \emph{not} undergo prompt black hole (BH) collapse.Finally, the presence of a  very fast kilonova ejecta component (Figure \ref{Fig:EkGammaBeta}) has important implications on still-open questions pertaining to the existence a free-neutron component of the ejecta possibly powering a short-lived luminous UV/optical transient, and the origin of subluminous gamma-rays produced in GRB\,170817A from the breakout of the cocoon shock from the merger ejecta.

Radiation powered by an energy release associated with the compact-object remnant in the form of accretion on a BH remnant 
offers an alternative explanation to the presence of an X-ray excess that is not accompanied by bright radio emission (\S\ref{SubSec:BHaccretion}). The detected X-ray luminosity $L_x\sim 5\times 10^{38}\rm{\,erg\,s^{-1}}$ is $\approx L_{\rm Edd}$  
for a compact-object with mass of a few $\rm{M_{\odot}}$. 
A long-lived NS cannot be entirely ruled out, but we conclude that it is an unlikely scenario based on the exceedingly low magnetic field $B\approx 10^{9}$\,G necessary to match the observed X-ray luminosity (\S\ref{SubSec:magnetar}).  
In analogy to stellar-mass compact-objects  accreting close to or above the Eddington rate, i.e. X-ray binaries (XRBs) in the ``soft'' state and ultra-luminous X-ray (ULXs) sources significant suppression of the radio emission can be expected.
Unlike the kilonova afterglow, where the radio emission is expected to brighten with time (Figure \ref{Fig:Xrays}), this accretion model predicts a constant or declining X-ray emission \emph{without} accompanying bright radio emission.

Observations of GW\,170817 are mapping an uncharted territory of the BNS merger phenomenology and have far-reaching theoretical implications. Measuring the time of peak of the kilonova afterglow, which probed the ejecta dynamics independent of shock microphysics,  would offer a unique opportunity to do calorimetry of the kilonova's fastest ejecta. Alternatively, the detection of a constant (or declining) source of X-ray emission in the next thousands of days that is not accompanied by bright radio emission will unveil how accretion processes work on a compact-object remnant of a BNS merger a few years after its birth.

\begin{deluxetable*}{cccccccc}
\tablecaption{\label{Tab:xrayanalysistab}  Observed and inferred properties of the X-ray counterpart of GW\,170817 as constrained by a spectral analysis of \emph{CXO} data with model \texttt{tbabs*ztbabs*cflux(pow)} within \emph{Xspec}. The net count-rate is computed for 1\arcsec\, region, using source and background counts from \texttt{ds9}. We adopted a Galactic neutral hydrogen column density in the direction of the transient of NH$_{\rm gal}=0.0784\times 10^{22}\,\rm{cm^{-2}}$ and no intrinsic absorption.  The uncertainties on the X-ray spectral parameters (photon index $\Gamma$ and unabsorbed 0.3- 10 keV flux) have been computed with MCMC sampling and are reported at the $1\,\sigma$ c.l.. Upper limits are reported at the $3\,\sigma$ c.l. {We provide two flux calibrations for the latest two epochs at $\delta t \sim 939$ and $1234$\,days for which the photon statistics is too limited to constrain the photon index: first we assume a photon index that is the best fitting value from the joint spectral fitting of all CXO observations $\Gamma = 1.603$  (see \S\ref{SubSec:CXOspecanalysis}). Second, we provide a flux calibration that assumes a spectral model consistent with a jet afterglow origin of the detected X-rays. From the posterior distribution of the $p$ parameter of models presented in \S\ref{Sec:Excess_stats}, we infer $\Gamma=1.565\pm 0.025$}.} 
\tabletypesize{\small}
\tablecolumns{7} 
\tablewidth{\textwidth}
\tablehead{\colhead{$\delta$t$^1$}  & \colhead{Significance$^2$} & \colhead{Exposure} & \colhead{Net count-rate$^3$} & \colhead{$\Gamma^4$}  & \colhead{Unabsorbed Flux} & \colhead{Luminosity$^5$} & \colhead{PI}\\
\colhead{(days)} & \colhead{($\sigma$)} & \colhead{(ks)} & \colhead{(10$^{-4}$\,ct/s)} & \colhead{} & \colhead{(10$^{-15}$\,erg\,cm$^{-2}$\,s$^{-1}$)} & \colhead{(10$^{38}$\,erg \,s$^{-1}$)} & \colhead{}\\
\colhead{} & \colhead{} & \colhead{} & \colhead{(0.5-8 keV)} & \colhead{} & \colhead{(0.3-10 keV)} & \colhead{(0.3-10 keV)}& \colhead{}}
\startdata
\rule{0pt}{3ex}$2.33^6$ & --  & $24.60$ & $<1.2$  & $1.4$  & $<1.9$  & $<3.75$ & Fong\\[2ex]
$9.19$ & $>8$ & $49.41$ &  $ 2.36 \pm 0.70$ & $0.78_{-0.56}^{+0.67}$ & $6.80_{-2.92}^{+2.82}$  & $13.5_{-5.79}^{+5.59}$ & Troja\\[2ex]
$15.39$ &  $>8$ & $96.1$ &  $2.95 \pm 0.56$ & $2.05_{-0.33}^{+0.49}$ & $5.32_{-0.99}^{+1.42}$  & $10.6_{-1.97}^{+2.81}$ & Haggard, Troja\\[2ex]
$108.39$ & $>8$ & $98.83$ & $13.5 \pm 1.17$ & $1.58_{-0.16}^{+016}$ & $25.6_{-2.34}^{+2.49}$ & $50.8_{-4.65}^{+4.93}$ & Wilkes \\[2ex]
$157.76$ & $>8$ & $104.85$  & $13.7 \pm 1.14$ & $1.64_{-0.18}^{+0.15}$ & $26.7_{-2.33}^{+2.90}$ & $52.8_{-4.63}^{+5.74}$ & Wilkes\\[2ex]
$259.67$ & $>8$ & $96.78$  & $6.85 \pm 0.85$ & $1.47_{-0.22}^{+0.23}$ & $13.9_{-2.01}^{+2.13}$ & $27.6_{-3.98}^{+4.22}$ & Wilkes\\[2ex]
$358.61$ & $>8$ & $67.16$  & $3.94 \pm 0.77$ & $2.02_{-0.34}^{+0.44}$ & $7.67_{-1.46}^{+1.76}$ & $15.2_{-2.89}^{+3.50}$ & Troja\\[2ex]
$581.82$ &  $>8$ & $98.76$  & $1.44 \pm 0.39$ & $1.19_{-0.61}^{+0.89}$ & $3.88_{-1.40}^{+1.97}$ & $7.68_{-2.77}^{+3.90}$ & Margutti\\[2ex]
$741.48$ & $6.5$ & $98.86$  & $1.03 \pm 0.34$ & $0.92_{-0.77}^{+0.91}$& $3.32_{-1.42}^{+1.75}$ & $6.58_{-2.81}^{+3.46}$ & Troja\\[2ex]
$939.31$ & $5.4$ & $96.60$  & $0.75 \pm 0.29$ & $1.603$ & $1.81_{-0.94}^{+0.79}$ & $3.59_{-1.86}^{+1.57}$ & Margutti\\[2ex] 
$1234.11$ & $7.2$ & $189.06$  & $0.77 \pm 0.21$ & $1.603$ & $2.31_{-0.81}^{+0.57}$ & $4.57_{-1.61}^{+1.13}$ & Margutti\\[2ex]
\hline
$939.31$ &  &   &  & $1.565 \pm 0.025$ & $2.14_{-1.35}^{+0.74}$ & $4.23_{-2.69}^{+1.46}$ & Margutti\\[2ex] 
$1234.11$ &  &   & & $1.565 \pm 0.025$ & $2.33_{-0.85}^{+0.60}$ & $4.62_{-1.69}^{+1.19}$ & Margutti\\[2ex]
\enddata
\smallskip
\footnotesize
$^1$ Exposure-time weighted average time since merger of all the observations within an epoch. The obsIDs within each epoch are as follows: 9\,days: 19294; 15\,days: 18988, 20728; 108\,days: 20860, 28061; 158\,days: 20936, 20937, 20938, 20939, and 20945; 260d: 21080, and 21090; 359\,days: 21371; 582\,days: 21322, 22157, and 22158; 742\,days: 21372, 22736, and 22737; 939\,days: 21323, 23183, 23184, and 23185; and 1234\,days: 22677, 24887, 24888, 24889,2 3870, 24923, and 24924. \\ $^2$ Gaussian equivalent.\\
$^3$ Inferred from \texttt{dmcopy} and energy filtering in channels 500-8000. \\$^4$Spectral photon index, where $F_{\rm \nu}\propto \nu^{-\beta}$ and $\Gamma=\beta+1$. \\ $^5$Calculated using a distance of 40.7 Mpc \citep{Cantiello+2018} \\
$^6$ From \cite{Margutti+2018binary}. 
\end{deluxetable*}

\begin{deluxetable*}{ccccccc}
\tablecaption{\label{tab:radioobslogtab}Radio Observations Log. Time on source for the VLA observations was calculated using the CASA analysis utilities task \texttt{timeOnSource}.}
\tabletypesize{\small}
\tablecolumns{7} 
\tablewidth{\textwidth}
\tablehead{\colhead{Start Date} & \colhead{$\delta$t} & \colhead{Observatory} & \colhead{Program/Project} & \colhead{On Source} & \colhead{Mean Frequency} & \colhead{Frequency Range} \\
        \colhead{UTC} & \colhead{(days)} & \colhead{} & \colhead{} & \colhead{Time (minutes)}  & \colhead{(GHz)} & \colhead{(GHz)}}
\startdata
        15\textsuperscript{th} Dec. 2020& 1216.08 & VLA & SL0449 & 204.23 & 3 & 2-4\\
        27\textsuperscript{th} Dec. 2020& 1228.02 & VLA & SL0449 & 204.23 & 3 & 2-4\\
        2\textsuperscript{nd} Feb. 2021& 1264.95 & VLA & SM0329 & 204.27 & 3 & 2-4\\
        10\textsuperscript{th} Feb. 2021& 1272.88 & VLA & SM0329 & 164.40 & 15 & 12-18 \\
        3\textsuperscript{rd} Jan. 2021 & 1234.66 & MeerKAT & DDT-20201218-JB-01 & 434.40 & 0.816 & 0.544-1.088 \\
\enddata
\end{deluxetable*}

\begin{deluxetable*}{ccccc}
\tablecaption{\label{tab:Pchance} { Chance probability of measuring a number of X-ray photon counts at least as extreme as the one observed at $t_1$, $t_2$ and combined, as a result of a stochastic fluctuation of the source and background. See \S\ref{Sec:Excess_stats} for details.} }
\tabletypesize{\small}
\tablecolumns{7} 
\tablewidth{\textwidth}
\tablehead{\colhead{$t_{start}$} & \colhead{$P_1$} & \colhead{$P_2$} & \colhead{$P_1\times P_2$} &  \colhead{$P_{combined}$} \\
        \colhead{(days)} &  & & }
\startdata
157.	&			6.0$\times 10^{-2}$ 	&		2.6$\times 10^{-4}$ &	1.7$\times 10^{-5}$& 9.5$\times 10^{-5}$\\
163.	&			1.1$\times 10^{-1}$		&		9.7$\times 10^{-4}$	&	1.2$\times 10^{-4}$& 6.7$\times 10^{-4}$\\
172.	&			1.2$\times 10^{-1}$ 	&		1.2$\times 10^{-3}$	&	1.8$\times 10^{-4}$& 9.5$\times 10^{-4}$ \\
196.	&			1.4$\times 10^{-1}$     &		1.5$\times 10^{-3}$	&	2.5$\times 10^{-4}$& 1.4$\times 10^{-4}$\\
209.	&			1.4$\times 10^{-1}$ 	&		1.5$\times 10^{-3}$	&	2.5$\times 10^{-4}$& 1.3$\times 10^{-4}$ \\
215.	&			1.3$\times 10^{-1}$  	&		1.5$\times 10^{-3}$	&	2.4$\times 10^{-4}$& 1.2$\times 10^{-4}$ \\
230.	&			1.4$\times 10^{-1}$ 	&		1.5$\times 10^{-3}$	&	2.4$\times 10^{-4}$& 1.3$\times 10^{-4}$ \\
\hline
$\gamma_B$ & $P_1$ & $P_2$ & $P_1\times P_2$ &  \colhead{$P_{combined}$} \\
\hline
7 &		1.5$\times 10^{-1}$	 &  2.8$\times 10^{-3}$  &	4.7$\times 10^{-4}$&	2.1$\times 10^{-3}$ 	\\
10 &	1.2$\times 10^{-1}$ & 2.7$\times 10^{-3}$ &  	3.4$\times 10^{-4}$& 1.4$\times 10^{-3}$	\\
12 &    7.2$\times 10^{-2}$ &	7.4$\times 10^{-4}$ &	5.6$\times 10^{-5}$&2.6$\times 10^{-4}$	\\ 
\enddata
\end{deluxetable*}

\section*{ACKNOWLEDGEMENTS}
{We thank the referees for their constructive input on the earlier draft of the manuscript. } A.~Hajela is partially supported by a Future Investigators in NASA Earth and Space Science and Technology (FINESST) award \#\,80NSSC19K1422.
This research was supported in part by the National Science Foundation under Grant No. AST-1909796 and AST-1944985, by NASA through Chandra Award Number G09-20058A and through Space Telescope Science Institute program \#15606. 
K.~D.~Alexander is supported by NASA through NASA Hubble
Fellowship grant \#HST-HF2-51403.001-A awarded by
the Space Telescope Science Institute, which is operated by the Association of Universities for Research in Astronomy, Inc., for NASA, under contract NAS5-26555.
B.~D.~Metzger is supported by NSF grant AST-2002577 and NASA grants 80NSSC20K0909, NNX17AK43G.
A.~Kathirgamaraju acknowledges support from the Gordon and Betty Moore Foundation through Grant GBMF5076.
D.~Radice acknowledges support from the U.S. Department of Energy, Office of Science, Division of Nuclear Physics under Award Number(s) DE-SC0021177 and from the National Science Foundation under Grant No. PHY-2011725.
S.~Bernuzzi acknowledges support by the EU H2020 under ERC Starting
 Grant, no.~BinGraSp-714626.
L. ~Sironi acknowledges support from the Sloan Fellowship, the Cottrell Scholars Award, NASA 80NSSC18K1104 and NSF PHY-1903412.
I. ~Heywood acknowledges support from the UK Science and Technology Facilities Council [ST/N000919/1] and the South African Radio Astronomy Observatory which is a facility of the National Research Foundation (NRF), an agency of the Department of Science and Innovation.
B.~Margalit is supported by NASA through NASA Hubble Fellowship grant No. HST-HF2-51412.001-A awarded by the Space Telescope Science Institute, which is operated by the Association of Universities for Research in Astronomy, Inc., for NASA under contract NAS5-26555.
R. ~Barniol Duran acknowledges support from National Science Foundation (NSF) under grant 1816694.
V.~A.~Villar is supported by the Simons Foundation through a Simons Junior Fellowship (\#718240). 
M.~Nicholl is supported by a Royal Astronomical Society Research Fellowship and by the European Research Council (ERC) under the European Union’s Horizon 2020 research and innovation programme (grant agreement No.~948381).
The Berger Time Domain group at Harvard is supported in part by NSF and NASA grants, as well as by the NSF under Cooperative Agreement PHY-2019786 (The NSFAI  Institute  for  Artificial  Intelligence  and  Fundamental  Interactions  http://iaifi.org/).   

The scientific results reported in this article are based to a significant degree on observations made by the \emph{Chandra} X-ray Observatory, and the data obtained from the Chandra Data Archive. Partial support for this work was provided by the National Aeronautics and Space Administration through Chandra Award Number GO1-22075X issued by the Chandra X-ray Center, which is operated by the Smithsonian Astrophysical Observatory for and on behalf of the National Aeronautics Space Administration under contract NAS8-03060.
The National Radio Astronomy Observatory is a facility of the National Science Foundation operated under cooperative agreement by Associated Universities, Inc.
The MeerKAT telescope is operated by the South African Radio Astronomy Observatory, which is a facility of the National Research Foundation, an agency of the Department of Science and Innovation.

\bibliography{ahbib2gw.bib}{}
\bibliographystyle{aasjournal}

\appendix
\section{X-ray Flux calibration}
\label{Appendix}
\begin{figure}[!t]
\begin{center}
\scalebox{1.}
{\includegraphics[width=\textwidth]{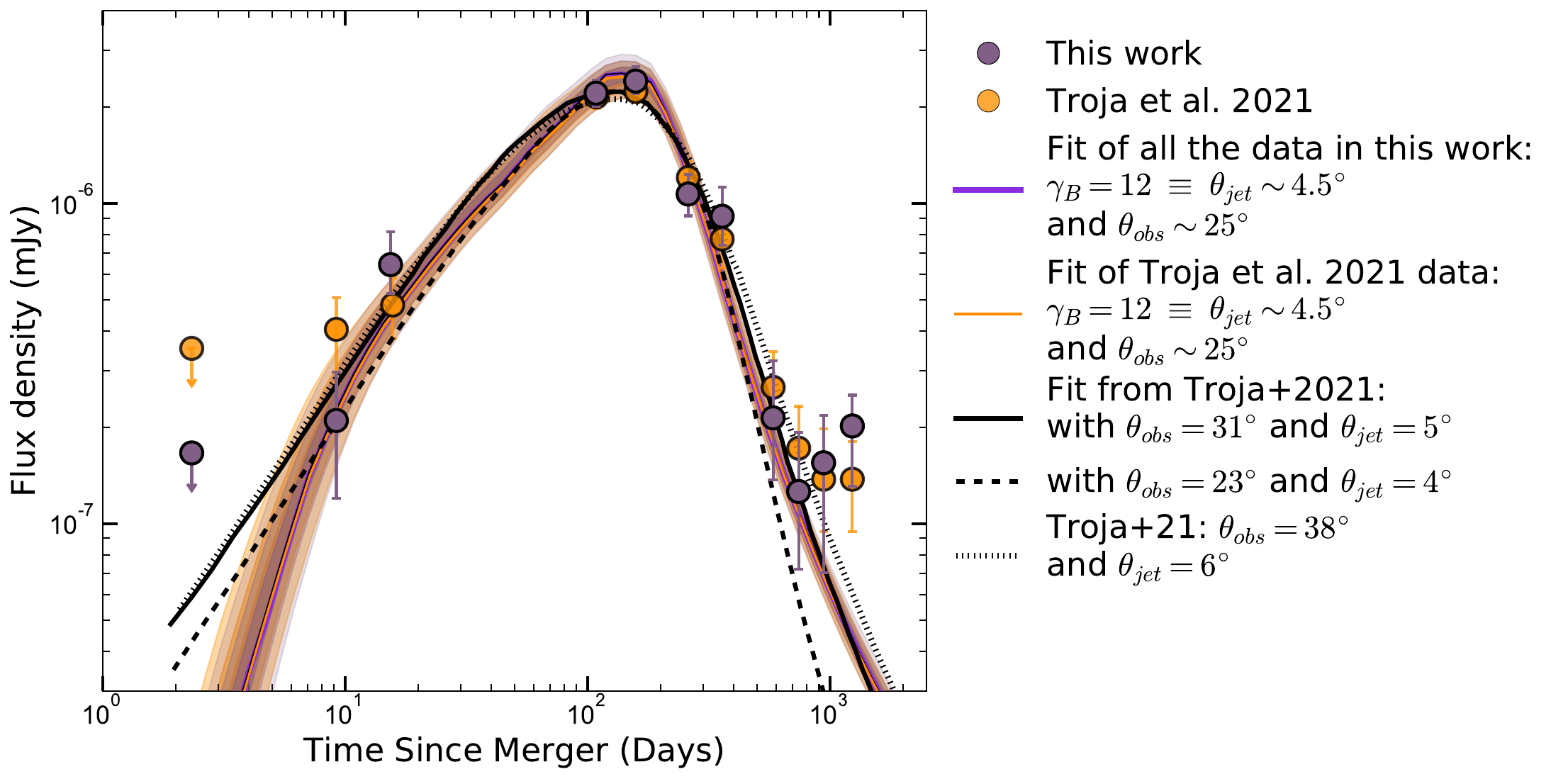}}
\caption{
Comparison of $1$\,keV flux densities derived in this work (purple circles, derived from the fluxes reported in Table \ref{Tab:xrayanalysistab}), where we do not assume a photon index based on the jet afterglow modeling, with those calculated by \cite{Troja+2021} (orange circles) where also the photon index is free for all epochs, as noted in their Table 1. We note that all the detections are consistent within $\lesssim 0.9\sigma$ uncertainties at all epochs. 
The colored bands are the 68\%, 97.5\%, and 99.8\% confidence interval of the fits obtained from fitting our data including the latest epochs at $\delta t > 900$\,days ( purple bands, same as in Figure \ref{Fig:JetfitLCcorner}) and from fitting all the data from \cite{Troja+2021} (in orange) using \texttt{JetFit}, with $\gamma_{\rm B} = 12$, $n = 0.01\,\rm{cm^{-3}}$, and $\epsilon_{\rm{e}}=0.1$ fixed. Even though these fits were obtained for all the flux densities derived using their respective spectral indices, for plotting purposes only, we have shown our $\delta t> 900$\,days flux densities (in purple) that were obtained using a fixed $\Gamma$. The best-fits derived in this work using \texttt{JetFit} with $\gamma_{\rm B} = 12$ and $\gamma_{\rm B} =10$ fixed as discussed in \S\ref{Sec:Excess_stats} are represented by solid and dashed purple lines, respectively. The best-fits obtained in \citealt{Troja+2021} are plotted in black lines. }
\label{Fig:xraycomparison}
\end{center}
\vspace{-0.5cm}
\end{figure}

In this Appendix we provide additional details on the comparison between the flux calibration of the X-ray data from this work and  the analysis of our data set by \cite{Troja+2021}. Our X-ray data analysis and the limitations of the data treatment by \cite{Troja+2021} are described in \S\ref{SubSec:CXOspecanalysis}. Figure \ref{Fig:xraycomparison} shows that the two flux calibrations lead to X-ray fluxes that are within $0.9\,\sigma$. There is thus no statistical tension between the two flux calibrations.  We further show the best fitting jet-afterglow  model that is used by \cite{Troja+2021} to compute the significance of the X-ray excess (black solid line, $\theta_{obs}=31^{\circ}$, $\theta_{jet}=5^{\circ}$), as well as the best-fitting model of the entire afterglow light-curve dataset (i.e. including the last two X-ray epochs) by \cite{Troja+2021}, which has $\theta_{obs}=38^{\circ}$, $\theta_{jet}=6^{\circ}$. This model is in tension with the inferences from the VLBI observations by \cite{Mooley+2018superluminal, Ghirlanda+2019}.  We present with a dashed black line the model by \cite{Ryan+2020} (also presented by \citealt{Troja+2021}, their Figure 5) that is consistent with the VLBI measurements, and that would lead to the inference of a larger discrepancy between the late-time X-ray observations and the model expectations.

As a proof of concept we have fitted the data at $\delta t < 1300$ days with \texttt{JetFit} using $\gamma_{\rm B}=12$. First, we included the X-ray fluxes from our data reduction where we leave the photon index as a free parameter for all epochs. Even though these fits were obtained for all the flux densities derived using their respective spectral indices, for plotting purposes only, we have shown our $\delta t> 900$\,days flux densities (purple points in Figure \ref{Fig:xraycomparison}) that were obtained using a fixed $\Gamma$. Second, we have repeated the same exercise by including the X-ray fluxes by \cite{Troja+2021}. Figure \ref{Fig:xraycomparison} shows complete overlap of the 68\% confidence regions of the two best-fitting models at all times.

\section{{\texttt{JetFit} corner plot}}
\label{Appendix2}
\begin{figure}[h]
\begin{center}
\scalebox{1.}
{
\includegraphics[scale=0.35,trim = {0 0.15cm 0 0}, clip]{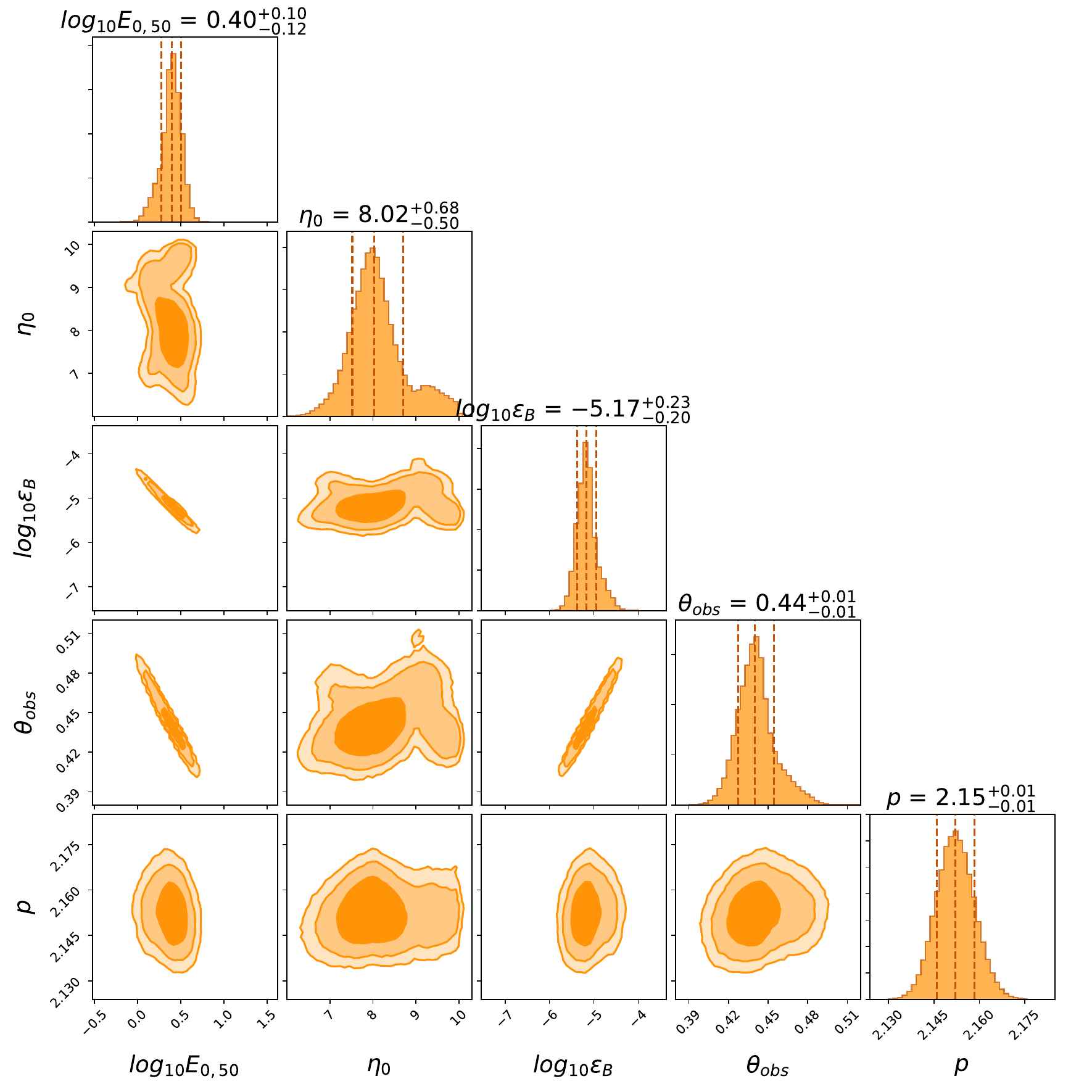}}
\caption{One- and two-dimensional projections of the posterior distributions of the model's free parameters. Vertical dashed lines mark the 16\textsuperscript{th}, 50\textsuperscript{th}, and 84\textsuperscript{th} percentiles of the marginalized distributions (i.e. the median and 1-$\sigma$ range). The contours are drawn at 68\%, 95\%, and 99\% credible levels. }
\label{Fig:postjetbreakmodelfit}
\end{center}
\vspace{-0.5cm}
\end{figure}
{In this paper, we have calculated the significance of a deviation of the observed X-rays using two different approaches - a jet-model dependent analysis, and a more universal, jet-model independent analysis. Both analyses independently result in an excess of X-ray emission with $\geq 3.1\sigma$ (Gaussian equivalent) confidence level. 
In Figure \ref{Fig:postjetbreakmodelfit}, we show the best-fitting universal post jet-break model. As mentioned in \S\ref{Sec:Excess_stats}, the fitting included the observations taken between $t_{\rm start} < t < 900$\,days. As seen from Figure \ref{Fig:postjetbreakmodelfit}, and the residual plot in the top panel of Figure \ref{Fig:JetfitLCcorner}, the fits start gradually diverging (although not to any significant level) from the data at $t > 300$\,days. This further strengthens our argument on the emergence of a new component of emission as it is likely to gradually manifest with time, as opposed to an abrupt appearance at some point in time, as the jet afterglow fades away.}
\end{document}